\documentclass[%
 reprint,
amsmath,amssymb,aps,prc,
]{revtex4-2}

\usepackage{graphicx}
\usepackage{dcolumn}
\usepackage{bm}
\usepackage{isotope}

\newcommand{\ice}[1]{\relax}

\begin{document}


\title{Inelastic Antideutron Interaction  with Nuclei}


\author{E.S. Golubeva}
\email[]{golubeva@inr.ru}
\affiliation{Institute for Nuclear Research, Russian Academy of Sciences, 117312 Moscow, Russia}


\date{\today}

\begin{abstract}
  We suggest to use the intranuclear cascade model (INC) for simulation of
  antideutron-nucleus inelastic interaction in the region of antideutron
  energies of $100$ MeV/n $\le T_{kin}\le 25-30$ GeV/n for a wide range of target nuclei. With the help of the model we obtain the cross-section
  $\sigma_{in}$ for inelastic antideutron-nucleus interaction and compare it
  with the ALICE experimental data.
 
The model is in good agreement with the available experimental data on the
inelastic interaction of antideutrons with tantalum nucleus.  We have created
a Monte Carlo generator for simulation of inelastic interaction of
antideutrons with nuclei. This generator can be useful in the preparation and
analysis of experiments to search for antideutrons near the Earth.
\end{abstract}


\maketitle

\section{Introduction}

The existence of antimatter follows from fundamental symmetries in particle
physics. There are no natural antinucleons and antinuclei on the Earth, they
can only be produced in the interactions of elementary particles and nuclei at
high energies. Although the antinucleon-nucleus interaction has been studied
in detail both experimentally and theoretically (see review \cite{Richard:2022tpn}), experimental
data on the interaction of light antinuclei with the matter are extremely
limited. There are several publications devoted to the study of the
interaction of antideutrons with nuclei of the matter performed at the IHEP
accelerator in Serpukhov \cite{Denisov:1971im,Binon:1970yu,Andreev:1990mk}. Recently, interest in the study of
antimatter has been renewed due to observation of light antinuclei
$\bar d$($\bar p\bar n$), $^3\bar He$($\bar p\bar p\bar n$) and
$^4\bar He$($\bar p\bar p\bar n\bar n$) in nucleon-nucleus and nucleus-nucleus
interactions at high energies at RHIC, SPS(CERN) accelerators and LHC(CERN)
\cite{E864:2000loc,STAR:2001pbk,PHENIX:2004vqi,NA49:2011blr,STAR:2019sjh,ALICE:2017nuf,ALICE:2019ikx,ALICE:2020chv,ALICE:2022xiu,ALICE:2020zhb}

Another important motivation for study lightest antinuclei is related with the dark matter problem.  By the 1970s, the existence of dark matter had been established and until now one of the most important tasks of modern physics is the experimental detection and identification of particles of cold dark matter (CDM). Current experiments focus on finding new particles  as constituents of the CDM.  Candidates for the role of such a particle  can be weakly interacting massive particles(WIMP) and it is hoped that dark matter can be found either by direct or indirect search. Annihilating pairs of WIMP’s could be indirectly detected with the help of gamma rays, positrons, and low-energy formation of antiprotons. Some models allow indirect detection of CDM via the registration of antideutrons \cite{Aramaki:2015pii,Donato:2008yx,Baer_2005,Bertone:2010zza}. Moreover, unlike indirect searches for dark matter using positrons, antiprotons, or gamma rays, which are complicated by the relatively high and uncertain astrophysical background, the search using antideuterons benefits from a suppressed natural cosmic-ray (CR) background. It was shown \cite{Baer_2005} that, for kinematic reasons, the various WIMP's annihilation processes under consideration lead to the generation of antideutrons with kinetic energy below 2-3 GeV/nucleon. For antideutrons resulting from CR interactions, the spectrum reaches a maximum at a much larger $T_{kin}$ ($\sim$10 GeV/nucleon). Thus, the low energy region is practically free from the background for CDM searching. This circumstance has lead to a surge of interest to the search for light antinuclei and to development of new experiments with satellites and balloons, such as AMS-02, GAPS, Alpha \cite{Ibarra:2012cc,Donato:1999gy,Kounine:2012ega,Hailey:2009fpa,Fornengo:2013osa}.

Recently, the AMS-02 detector on the International Space Station has detected
about 10 antihelium events. The AMS-02 data also include several events with
mass identification most consistent with antideuterons, though it is difficult
to eliminate the possibility that these mass measurements constitute the tail
of the antiproton distribution \cite{luque2024cosmicraypropagationmodelselucidate}.

Antinuclei are sensitive probes of WIMP's in the search of dark matter. If they exist, WIMP's should accumulate in the center of our galaxy, where they can annihilate into antinuclei. To calculate the expected flow of antinuclei near the Earth, it is necessary to know the probabilities of formation and annihilation of antinuclei in the Galaxy. For experiments on colliders, this essentially poses three problems:

 First, to understand the processes of antinuclei formation in order to simulate their appearance during the annihilation of dark matter and subsequent decays.
 
 Secondly, to quantify the formation of antinuclei in background reactions during collisions of energetic cosmic rays. 
 
 Third, to study the interaction of antinuclei with particles of matter in order to determine the transparency of the galaxy.

 To understand the propagation and interaction of antimatter in the interstellar medium, information about the inelastic interaction cross sections of the light antinuclei with nuclei is of crucial importance. Recently, the results of measurement of inelastic cross sections for antideutron interaction with the nuclei of ALICE detectors (CERN, LHC) were obtained in the momentum range 0.3$\le$ P $\le$ 4 GeV/c \cite{ALICE:2020zhb}. In the experiment, the antiparticle-to-particle yield ratios are compared with the detailed results simulated with GEANT4 for propagation of (anti)particles (this means that what is said applies to both the particle and the antiparticle) through the detector materials. The parameterizations of inelastic cross sections obtained within the Glauber approach \cite{Uzhinsky:2011zz,Galoyan:2016huf} are employed to simulate the passage of (anti)deutrons through the detector substance in GEANT4 \cite{Allison:2016lfl}.  In the Glauber model, the amplitudes obtained from independent experiments are utilized in expressions for cross sections, and, strictly speaking, there are no fitting parameters. However, in reality, additional parameterizations are needed for the radius of the target nucleus \cite{Uzhinsky:2011zz}. This approach leads to a good agreement with the experimental cross sections available for the antiproton-nucleus interaction, and predicts the cross sections for the interaction of the lightest antinuclei $\bar d$, $\bar t$, $^3\bar He$ and $^4\bar He$.

 Another possible way to determine inelastic antinucleus-nucleus cross sections is the INC model, which practically does not contain free fitting parameters. While the Glauber model allows one to calculate only the integral cross sections $\sigma_{tot}$, $\sigma_{in}$, the INC allows one to obtain information about a variety of interaction channels and gives a complete exclusive description of complex nuclear reactions, such as the interaction of antiprotons or antideutrons with nuclei. The first steps in studying the interaction of antideutrons with the tantalum nucleus at the momentum 12.2 GeV/c, and comparing the simulation results and experimental data were made in \cite{Andreev:1990mk}. This paper presents the first results of application the INC model to determine the cross-section of the antideutron-nucleus inelastic interaction. Section 2 provides a brief description of the INC model used to simulate the inelastic interaction of antideutrons with nuclei and determine cross-sections. Section 3 demonstrates the cross sections of inelastic interaction for (anti)protons and (anti)deuterons obtained within the framework of the INC model, compared with experimental data. Section 4 presents a comparison of the simulation results of the interaction of antideutrons with the tantalum nucleus with experimental data. Some predictions done  for the inelastic interaction of antideutrons with other nuclei. Our conclusions are formulated in Section 5.

 \section{Description of the model}

Inelastic nuclear reactions have a pronounced statistical character, since a large number of possible final states can be realized. The INC is a quasi-classical approximation and implements a statistical approach to the description of inelastic nuclear reactions. The statistical approach is the main one for such complex multi particle systems, when the description of the evolution of the wave function of the system is replaced by a description of the ensemble of its possible states. INC has long been successfully used to simulate nuclear reactions at intermediate energies of 50-100 MeV/n $\le T_{kin} \le$ 25-30 GeV/n. Using the method of successive collisions, the INC allows to trace the fate of all particles involved and born in the hadron-hadron interactions, taking into account the influence of the Fermi motion of the intranuclear nucleons. The fundamentals of the statistical approach to the description of high-energy particles and nuclei interaction with nuclei are described in detail in the monograph \cite{Barashenkov:1972zza}. 

The INC model of interaction of antinucleons with nuclei is used as the basis for description the intranuclear cascade initiated by an antideutron.  Approximations of experimental data on the $N N-$, $\pi N-$ and $\bar N N-$ collisions in vacuum are used to simulate elementary interactions in the nucleus.  Highly excited residual nuclei formed at the end of the intranuclear cascade subsequently decay. Along with the evaporation mechanism of disintegration of residual nuclei at moderate ($\le$  5 MeV/n)  excitation, the explosive mechanism of disintegration at high excitation ($\sim$ 7 MeV/n) is taken in to account. The detailed description  of the INC model of inelastic  $\bar N A-$ interaction is given in \cite{Golubeva:1988de}.

Let us take a closer look at some details of the model presented in this paper. The target nucleus is considered as a degenerate Fermi gas of free nucleons enclosed in a spherical potential well with a diffuse layer. The model takes into account the nonlinear effect of local decrease in the nuclear density in the process of development of intranuclear cascade \cite{Barashenkov:1971vh}. This is important in the case of multiple pions production in the antinucleon annihilation, especially for light target nuclei, when the condition of applicability of the cascade model $N_c \ll A_t$ is violated ($N_c$ is the number of cascade particles, $A_t$ is the number of nucleons in the target nucleus). This effect is even more important in the case of possible annihilation of both antinucleons in the nucleus.  During the development of the intranuclear cascade, more and more target nucleons are involved. Intranuclear nucleon becomes a cascade particle after the interaction. So, fast cascading particles knock out the intranuclear nucleons while  slower cascading particles move in the regions with a lower density of nuclear matter. The approach is described in detail in \cite{Golubeva:2018mrz}. This approximation works well for the high-density central regions of the nucleus, but is not entirely correct for the low-density nuclear periphery where the first antinucleon interaction occurs. The effect of a local decrease of nuclear density is important when a multiple cascading particles pass through the nucleus, but it does not play a role in the first interaction of the incoming particle with the nucleus. Our suggestion is to combine the standard cascade model with a continuous density distribution in the first interaction and a model, that takes into account the effect of density reduction in the subsequent development of the cascade. This approach makes it possible to correctly describe both the interaction of the initial particle in the region of the nuclear periphery and the passage of multiple cascading particles through the nucleus. It allows to obtain simultaneously the correct values of the inelastic cross-section determined by the first interaction, and the exclusive characteristics of such a complex process as antideutron-nucleus interaction.

 By analogy with the deuteron-nucleus interaction \cite{Barashenkov:1973ew}, the inelastic interaction of an antideutron is reduced to intranuclear cascades caused by its antinucleons in the target nucleus \cite{Andreev:1990mk}. Cascades generated by the antinucleons of the incoming antideutron turn out to be effectively connected due to the local decrease in the nuclear density of the target. In the model (anti)deutron is considered as a loosely coupled system – a "dumb-bell" of (anti)proton and (anti)neutron with the fixed distance $l=2R_d = 4,32*10^{-13}$ cm  between them. The direction of the axis of such a "dumb-bell" is isotopically distributed in space and does not change during the movement. Such a representation of the flying (anti)deuteron makes it possible to simulate the stripping reaction, when one (anti)nucleon of (anti)deutron  interacts with the target nucleus, and the second one continues to move without interaction.

  In the coordinate system of the deuteron, the distribution over the relative momenta of nucleons determined by the expression:
 \begin{equation}\label{momdistr}
 W_{d}(P) \sim \frac{ P^2 }{(P^{2}+m_{N}\epsilon)^2}\mbox{ ,}
\end{equation} 
where $\epsilon=0.00223$ $GeV/c^2$, $m_N$  and $P$ stand for the nucleon mass (in $GeV$) and its momentum (in $GeV/c$) respectively. This expression  represents the  square of the Fourier components of the approximate deuteron wave function \cite{Andreev:1990mk}.  It was shown in \cite{Barashenkov:1972zza} that the details
 of the $W_{d}(P)$ distribution have only a minor impact on the simulation results. We think that the above simple model for antideutron is quite justified in the first approximation.

Within the framework of the proposed model, the inelastic cross section of the hadron-nucleus interaction is defined as
\begin{equation}\label{sig1}
\sigma_{in}=\pi (r_{nucl}+\lambda/2)^{2}*{N_{in}/N_{tot}}\mbox{ ,}
\end{equation} 
where $r_{nucl} = r_{0}*A^{1⁄3}$ is the radius of the target nucleus, $\lambda $ is the wavelength of the incoming particle, which approximately takes into account the smearing of its trajectory. Thus, hadrons can interact with the nucleus if their centers move at a distance not exceeding $r_{int} =r_{nucl} +\lambda/2$ from the center of the nucleus.  The ratio of the number of inelastic interactions to the total number of interactions $N_{in}/N_{tot}$  is determined directly from the simulation.This ratio is sensitive to the interaction radius and the mean free path of initial (anti)nucleon in the target nucleus. Since the determination of the mean free path is based on experimental values of the total elementary interaction cross section of (anti)nucleon and does not contain additional parameters, the interaction radius $r_{int}$ can be considered as the only parameter of the model.
Further, an account of the size of (anti)deutron leads to the following form of the inelastic interaction cross section of  (anti)deuterons: 
\begin{equation}\label{sig2}
\sigma_{in}=\pi (r_{nucl}+R_{d}+\lambda/2)^{2})*{N_{in}/N_{tot}}\mbox{ ,}
\end{equation} where $R_{d} = 2,16$ fm is the average “radius” of the (anti)deuteron and $\lambda$ refers to the wavelength of the incoming (anti)deutron. 

The stripping reaction is considered as a special case of inelastic interaction. Thus, the problem is reduced to modeling of intranuclear cascade initiated by two (anti)nucleons  of (anti)deuteron or one of them in the case of stripping.

\section{Cross sections of inelastic (anti)nucleon-nucleus and (anti)deutron-nucleus interaction }

The INC model has been successfully used for a long time to simulate nuclear reactions, but it can also be applied to determine the cross-sections of inelastic interactions with nuclei. Before considering cross sections of (anti)deutron-nucleus interaction, let us analyze how the proposed model describes the experimental cross sections of inelastic (anti)proton–nucleus interaction.   Figures 1 and 2 show cross sections of inelastic interactions of protons and antiprotons with carbon and lead nuclei in the energy range of the incident particle from ~ 0.180 to 29 GeV. The experimental values are taken from compilation of experimental data for interactions of hadrons and nuclei with nuclei \cite{Barashenkov:1993zza}. With the exception of the lowest energy values of ~ 0.180 GeV, the calculation results satisfactorily describe the energy dependence of inelastic cross sections, both for protons and antiprotons, for light and heavy nuclei, although additional comparisons with experimental data are needed, including the cross sections of interaction for antineutrons. Based on the results of such comparisons, the determination of the the parameter $r_{int}$ can be refined.

\begin{figure}
\centering
\includegraphics[width=0.950\linewidth]{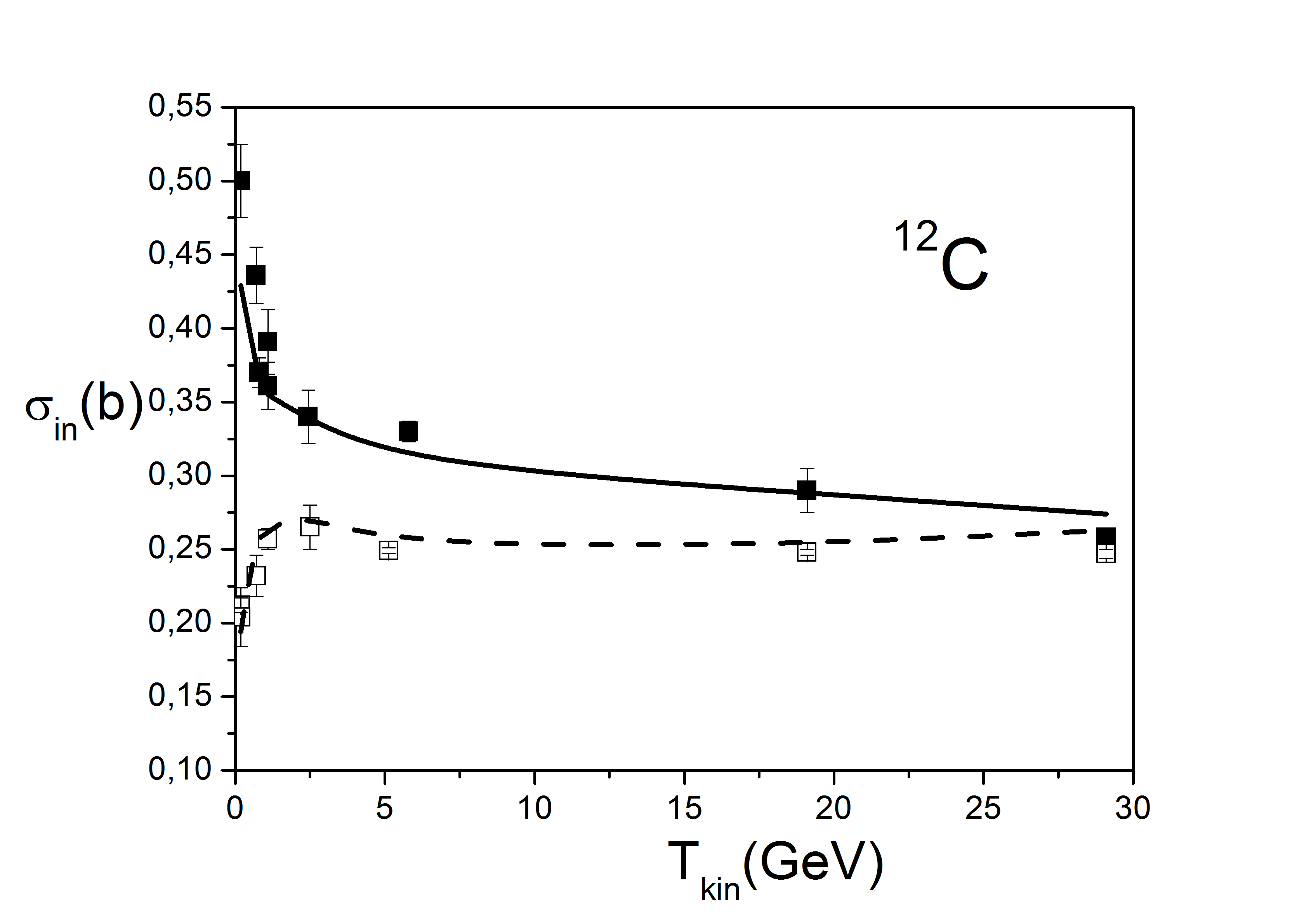}
\caption{Cross sections of inelastic interaction of protons and antiprotons
  with the nucleus $^{12}C$. Experimental values \cite{Barashenkov:1993zza}: the black squares:
  $\bar {p} +^{12}C$, the open squares: $p+^{12}C$.  The solid line is the
  calculation of $\bar {p} +^{12}C$.  The dotted line – calculation of
  $p+^{12}C$.}\label{fig_1}
\end{figure}

\begin{figure}
\centering
\includegraphics[width=0.950\linewidth]{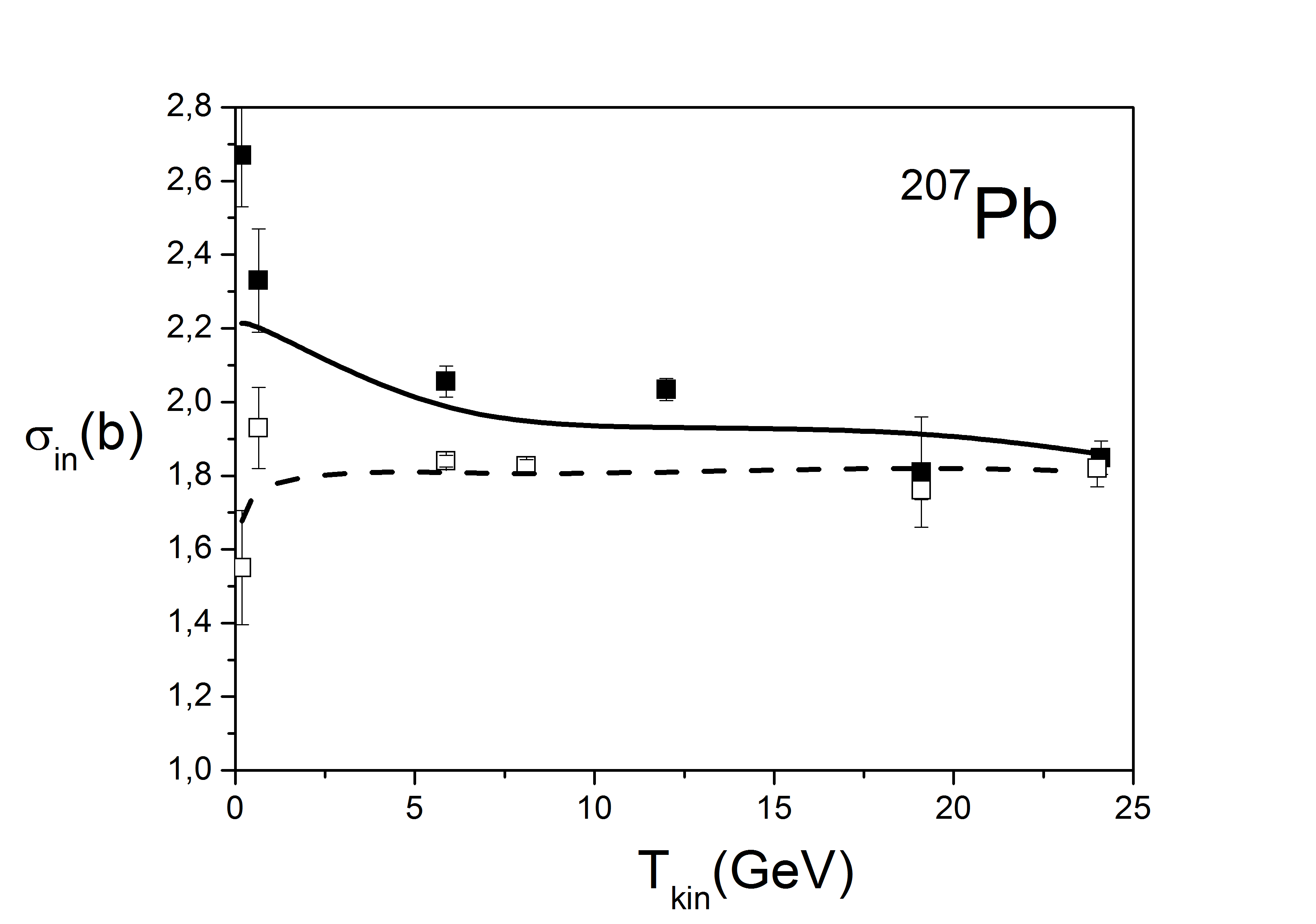}
\caption{Cross sections of inelastic interaction of protons and antiprotons with the $Pb$ nucleus. Experimental values \cite{Barashenkov:1993zza}: the black squares: $\bar {p} +Pb$, the open squares:  $p+Pb$.  The solid line is the calculation of $\bar {p} +Pb$. The dotted line – calculation of $p+Pb$.}\label{fig_2}
\end{figure}

Now let us compare the cross section of the inelastic interaction of deutrons with nuclei. Figure 3 shows the A-dependence of the inelastic interaction cross section of deuterons with nuclei at the energy of 0.160 GeV. The cross sections $\sigma_{in}$ obtained in the calculation include the stripping cross section $\sigma_{st}$. The calculation slightly underestimates the values of the inelastic cross section $\sigma_{in}$ for heavy nuclei. It should be noted here that at this energy the cross sections of diffraction as well as the Coulomb splitting of a deuteron should be added to the calculated  $\sigma_{in}$. This is a separate  task.   Contributions to the cross-section from these processes are not taken into account in this work.

\begin{figure}
\centering
\includegraphics[width=0.950\linewidth]{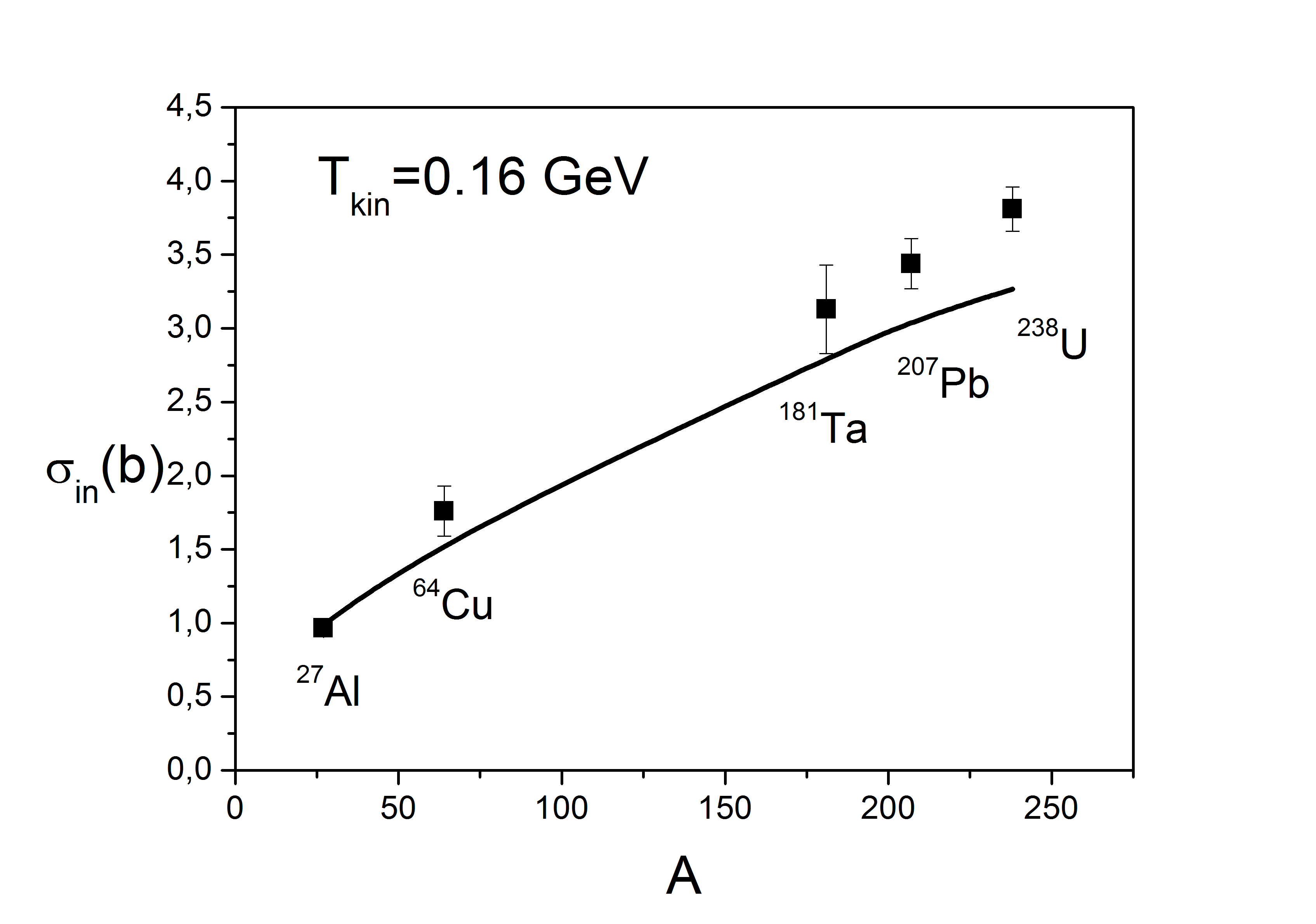}
\caption{A - dependence of the cross section of the inelastic interaction of deuterons with nuclei at an energy of 0.16 GeV in barns. The points are experimental data \cite{Barashenkov:1972zza}. The line is a calculation.}\label{fig_3}
\end{figure}

The agreement of the calculation results with experimental data strongly depends on the reproduction of experimental data selection during modeling. Figure 4 shows the experimental and calculated cross section of proton stripping $\sigma_{st}$, when the cross-section is the sum of proton stripping cross-section  and the cross-section of the process in which the neutron from the deutron goes through the target nucleus without interaction, and the proton after collision with the target nucleus flies out in a direction close to the original direction of motion of the deuteron. Although the conditions for the selection of experimental events in the simulation are reproduced fairly approximately, the theoretical and experimental cross sections $\sigma_{st}$ are in satisfactory agreement with each other.

\begin{figure}
\centering
\includegraphics[width=0.950\linewidth]{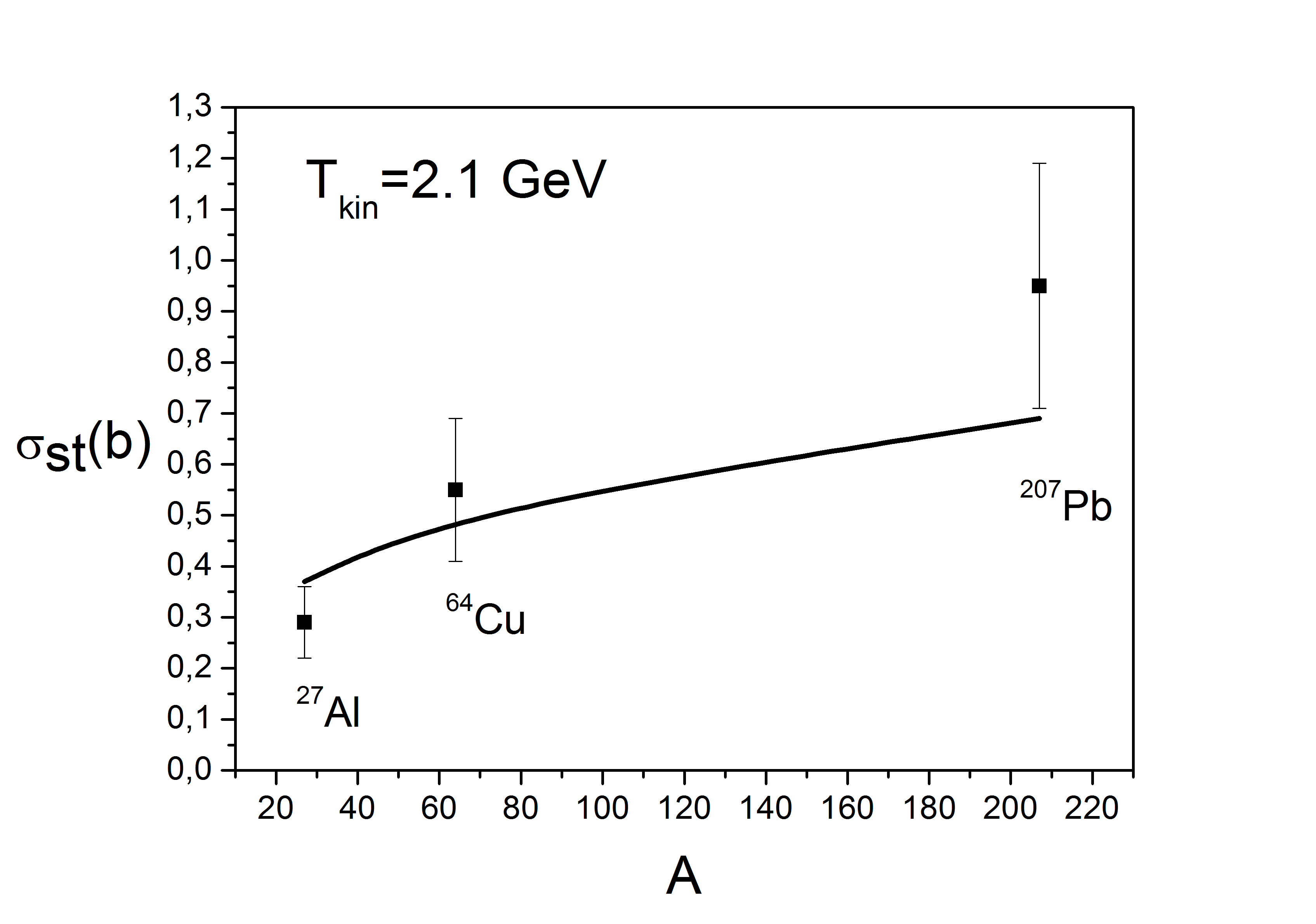}
\caption{A - dependence of the stripping cross section $\sigma_{st}$(see text) of the interaction of deuterons with nuclei at an energy of 2.1 GeV in barns.The points are experimental data \cite{Barashenkov:1972zza}. The line is a calculation.}\label{fig_4}
\end{figure}

\begin{figure}
\centering
\includegraphics[width=0.950\linewidth]{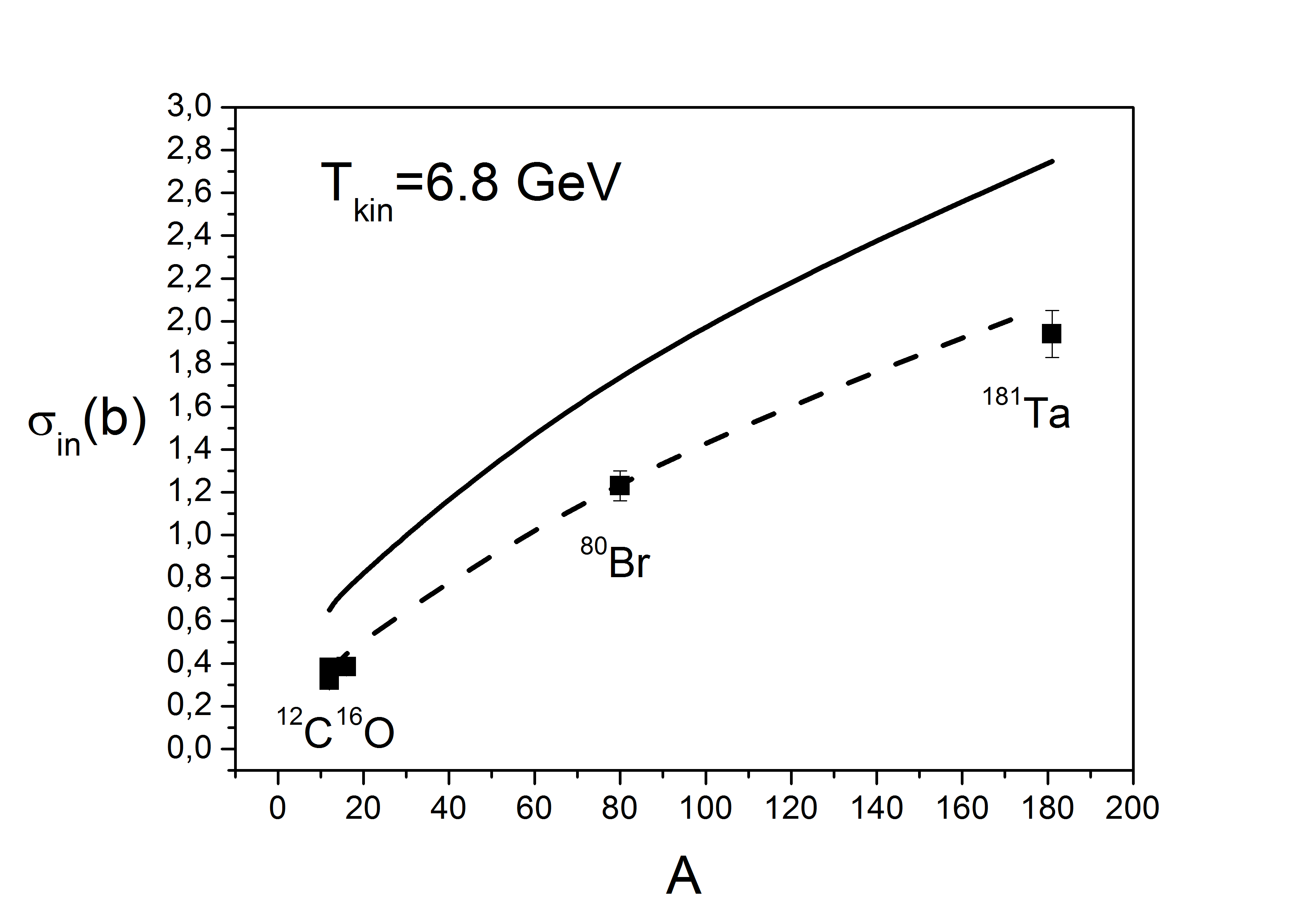}
\caption{A - dependence of deutron - nucleus cross section  $\sigma_{in}$ at the energy  of 6.8 GeV in barn. The points are experimental data \cite{Barashenkov:1993zza}. The solid line – calculation $\sigma_{in}$. The dashed line – calculation $\sigma_{in}-\sigma_{st}$.}\label{fig_5}
\end{figure}

Figure 5 shows the deuteron inelastic cross sections  with nuclei at the energy 6.8 GeV.  The solid curve shows the calculated values $\sigma_{in}$ and significantly exceeds the experimental values. This looks somewhat strange, since it follows from Figure 3 that the calculation gives rather understated cross sections $\sigma_{in}$. However, as we saw in Figure 4, it is very important in modeling to reproduce the experimental conditions for selecting events, which are not always clear from the description of the data. Figure 5 shows the values $\sigma_{in}-\sigma_{st}$ (dashed curve), where $\sigma_{st}$ is the cross section of proton stripping, and this gives a good agreement with the experimental data.

Thus, it can be said that in the case of deuteron-nucleus interaction, it is critically important to take into account the criteria for selecting events in the experiment and isolate the stripping reaction. For a more precise adjustment of the parameter $r_{int}$, it is necessary to rely on experimental data on the cross sections of inelastic deuteron-nucleus interactions, obtained with clear criteria for selecting events in a wide range of energies for light and heavy nuclei.

The first experimental data on the cross sections of inelastic interaction of antideutrons with nuclei were obtained at the IHEP accelerator in Serpukhov \cite{Denisov:1971im,Binon:1970yu}. The cross section $\sigma_{in}^{'}=\sigma_{in}-\sigma_{st}^{\bar{p}}$ was measured in the experiment. The  experimental values of $\sigma_{in}$ presented in Tables 1 and 2 include the calculated corrections $\sigma_{st}^{\bar{p}}$\cite{Binon:1970yu}.

Figures 6 and 7 show the experimental cross sections $\sigma_{in}$ (with the calculated correction  $\sigma_{st}^{\bar{p}}$), as well as the theoretical cross sections $\sigma_{in}$ (solid line) and $\sigma_{in}-\sigma_{st}^{\bar{p}}$ (dotted line). In general, there is satisfactory agreement between experimental and theoretical values, although the issue related to the contribution of stripping processes requires a more detailed analysis.

\begin{figure}
\centering
\includegraphics[width=0.950\linewidth]{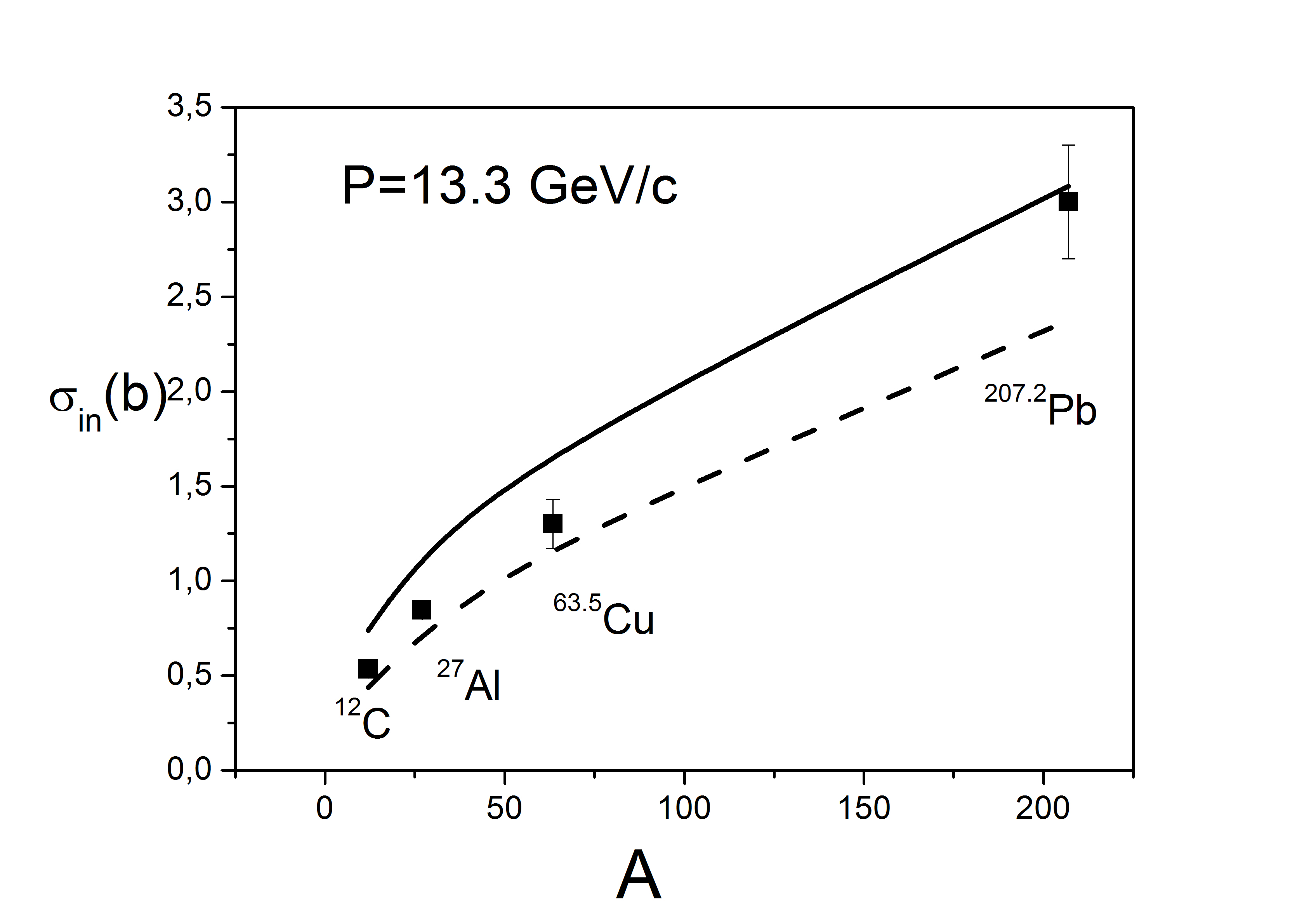}
\caption{A - dependence of antideutron - nucleus cross section  $\sigma_{in}$ at the momentum 13.3 GeV/c in barns. The points are experimental data. The solid line is the calculation of $\sigma_{in}$, the dashed line is the calculation of $ \sigma_{in}-\sigma_{st}^{\bar{p}}$. }\label{fig_6}
\end{figure}

\begin{table}
\centering
\begin{tabular}{|c|c|c|c|}
\hline
Target&$\sigma_{in}exp$&$\sigma_{in}calc$&$\sigma_{in}-\sigma_{st}^{\bar{p}}$\\
\hline
$^{12}C$&$0.437\pm{0.027}$& $0.739$&$0.437$ \\
\hline
$^{27}Al$&$0.845\pm{0.043}$& $1.144$&$0.727$ \\
\hline
$^{63.5}Cu$&$1.300\pm{0.130}$& $1.717$&$1.213$ \\
\hline
$^{207.2}Pb$&$3.000\pm{0.300}$& $3.084$&$2.375$ \\
\hline
\end{tabular}
\caption{\label{tab:widgets}Inelastic antideuteron - nucleus interaction cross sections  at P= 13.3 GeV/c in barn [2].}
\end{table}
 
\begin{figure}
\centering
\includegraphics[width=0.950\linewidth]{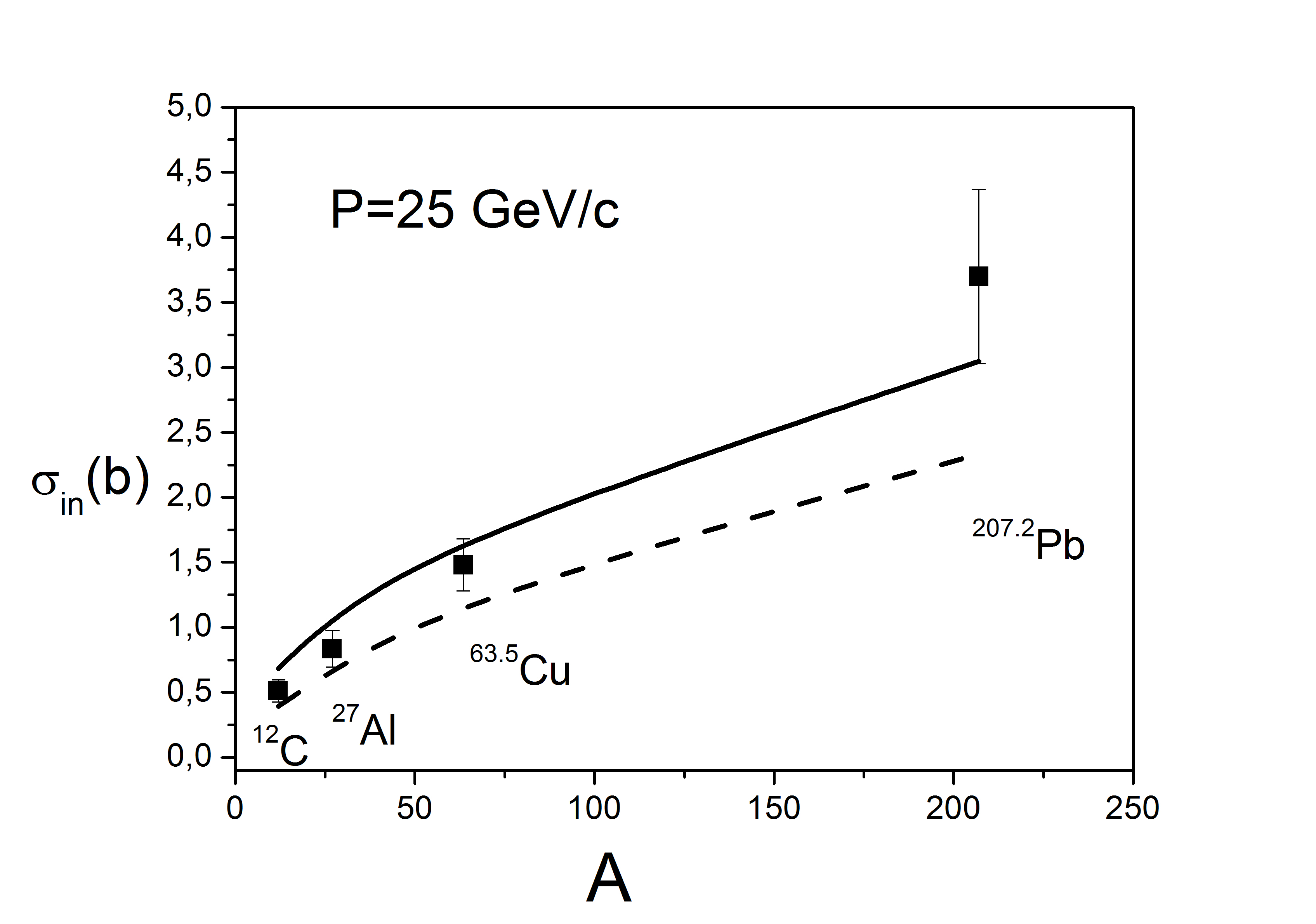}
\caption{A - dependence of antideutron - nucleus cross section  $\sigma_{in}$ at the momentum 25 GeV/c in barns.  The points are experimental data. The solid line is the calculation of $\sigma_{in}$, the dashed line is the calculation of $ \sigma_{in}-\sigma_{st}^{\bar{p}}$. }\label{fig_7}
\end{figure}

\begin{table}
\centering
\begin{tabular}{|c|c|c|c|}
\hline
Target&$\sigma_{in}exp$&$\sigma_{in}calc$&$\sigma_{in}-\sigma_{st}^{\bar{p}}$\\
\hline
$^{12}C$&$0.510\pm{0.085}$& $0.682$&$0.393$ \\
\hline
$^{27}Al$&$0.835\pm{0.140}$& $1.085$&$0.680$ \\
\hline
$^{63.5}Cu$&$1.480\pm{0.200}$& $1.712$&$1.223$ \\
\hline
$^{207.2}Pb$&$3.700\pm{0.670}$& $3.045$&$2.334$ \\
\hline
\end{tabular}
\caption{\label{tab:widgets}Inelastic antideuteron - nucleus interaction cross sections  at P= 25 GeV/c in barn [2].}
\end{table}

In fact, the cross sections of inelastic antideutron-nucleus interaction remained poorly studied except the data at high momentum, obtained in the 1970s  \cite{Denisov:1971im,Binon:1970yu}. Recent efforts of  ALICE collaboration have been aimed at studying of $\sigma_{in}$ of light antinuclei inelastic interaction with the nuclei of ALICE detectors  at the momentum range $0.3\leq p\leq 4$ GeV/c.  In the experiment antideutrons are formed during Pb collisions with energy in the CMS $\sqrt{S_{NN}}=5.02$ TeV per nucleon pair\cite{ALICE:2020zhb} and the ALICE detector itself is a target for interaction of (anti)protons and (anti)deuterons. The ratio of antiparticle  to particles yield obtained as a result of measurements was compared with the results of  the ALICE detector simulation, based on the GEANT4 version, which includes parametrization of the Glauber model for anti(deuteron) cross sections. The analysis of the antiproton/proton ratio serves as a starting point for this study, since the cross sections of inelastic interaction for anti(protons) are well defined experimentally.  The materials of the detector subsystems serve as a target for interaction of anti(deuterons). The ALICE detector material from the primary IP (interaction point)  to TOF (Time-Of-Flight) detector has an average atomic number $<Z>=14.8$ and a mass number $<A>=31.8$. For the detector material up to the middle of the TPC (Time Projection Chamber), an average atomic number is $<Z>$=8.5 and a mass number is $<A>=17.4$. These values were obtained by weighing the contribution of various materials with their density multiplied by the length intersected by the particles \cite{ALICE:2020zhb}.

\begin{figure}
\centering
\includegraphics[width=0.950\linewidth]{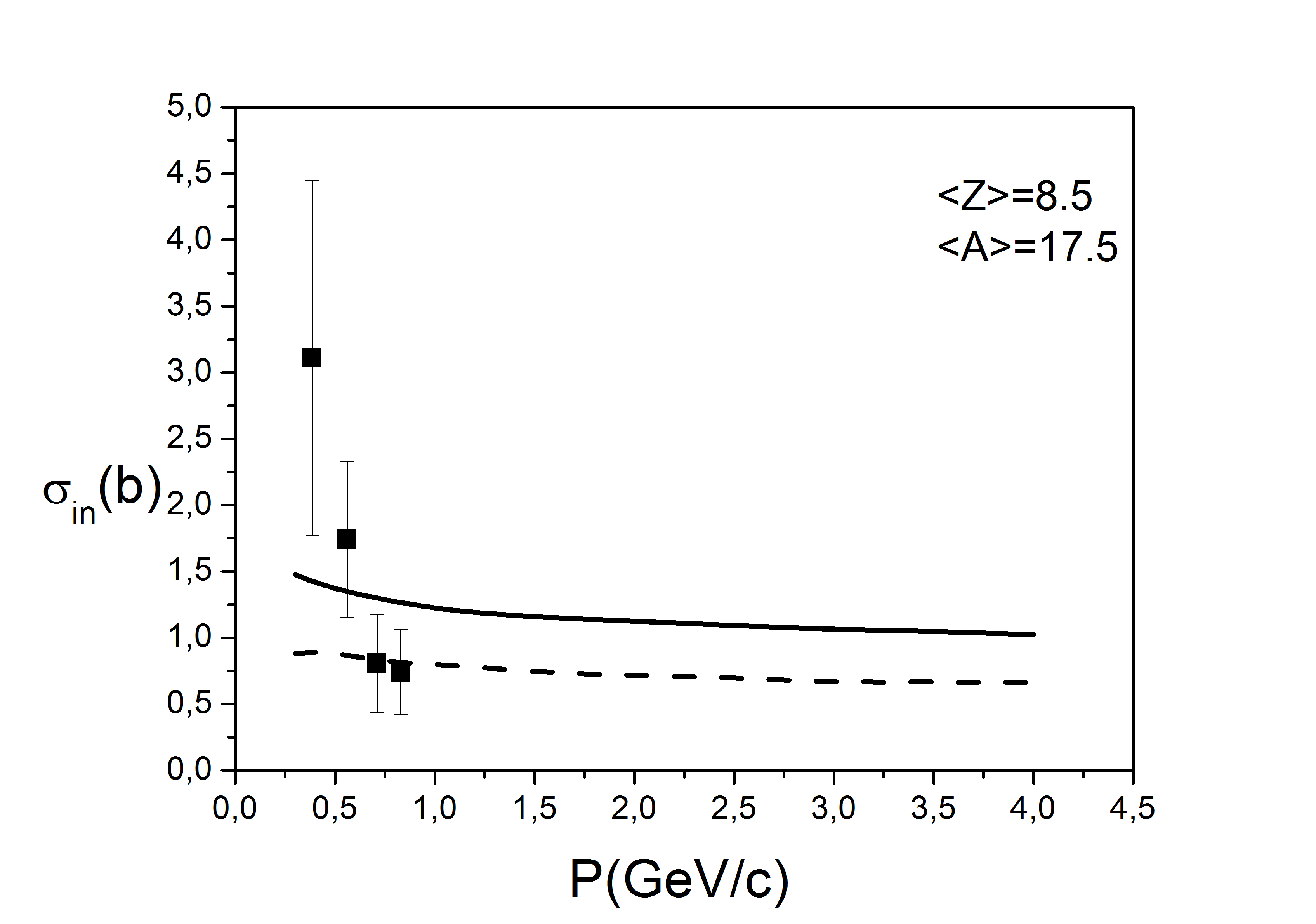}
\caption{Cross sections (in barns) of antideutrons inelastic interaction with an average light nucleus $<Z>$=8.5, $<A>$=17.5 of  ALICE detector.  The solid line is $\sigma_{in}$, the dashed line $\sigma_{in}-\sigma_{st}^{\bar{p}}$. The points are experimental data  \cite{ALICE:2020zhb}. }\label{fig_8}
\end{figure}  

\begin{figure}
\centering
\includegraphics[width=0.950\linewidth]{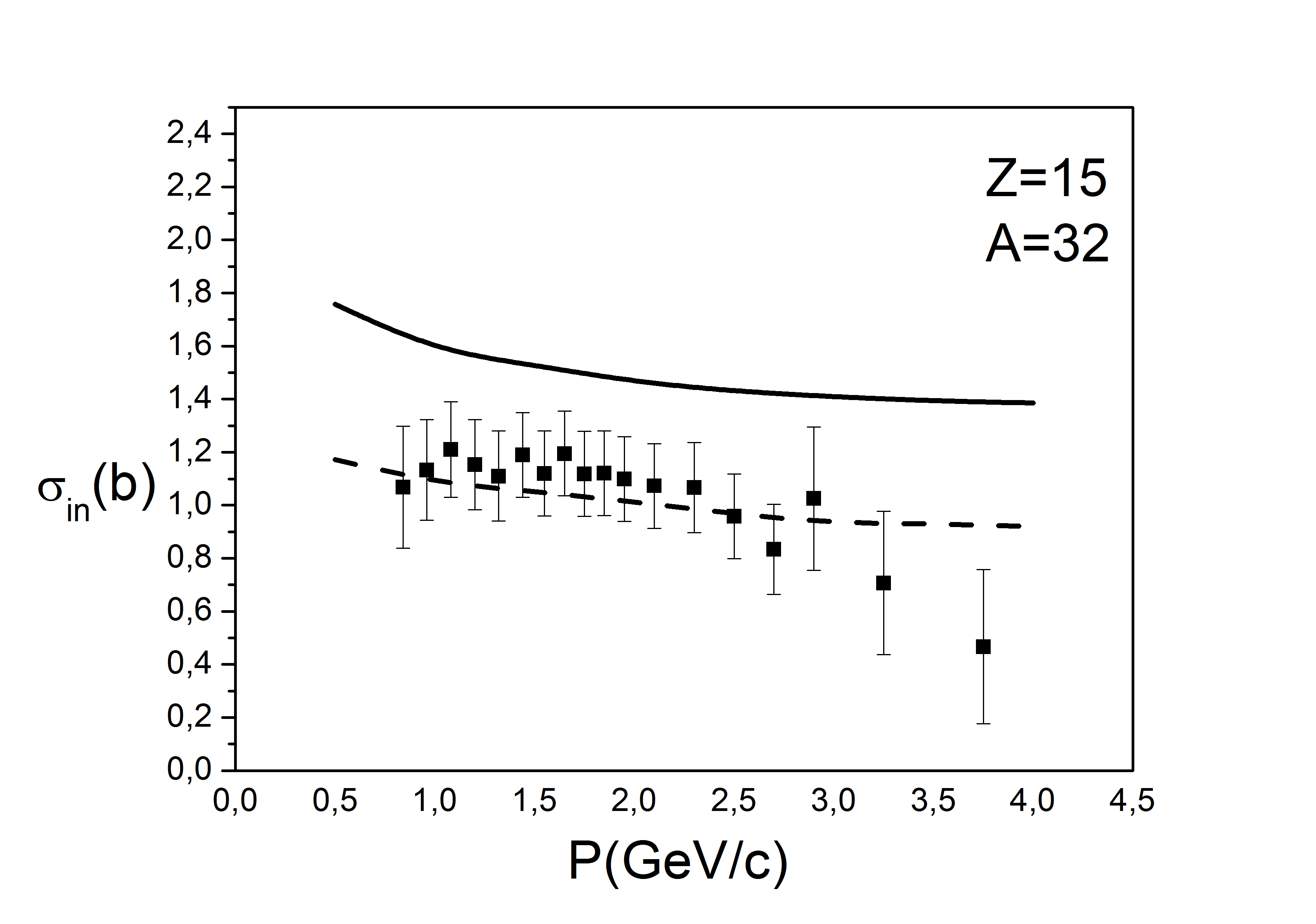}
\caption{Cross sections (in barns) of antideutrons inelastic interaction with an average nucleus Z=15, A=32 of  ALICE detector.  The solid line is $\sigma_{in}$, the dashed line $\sigma_{in}-\sigma_{st}^{\bar{p}}$. The points are experimental data \cite{ALICE:2020zhb}. }\label{fig_9}
\end{figure}

In present paper, as a first approximation, the simulation was carried out for the nucleus Z=15, A=32 of the first part and averaged over the nuclei Z=8, A=17 and Z=9, A=18 for the second part of the detector. Although it is necessary to check how much the cross section obtained by averaging calculations on all nuclei of the detector differs from the cross section calculated on the average nucleus. Figures 8 and 9 show the interaction cross sections of antideutrons with the averaged nuclei of the ALICE detector. The solid line corresponds to the inelastic cross section $\sigma_{in}$, and the dashed one to $ \sigma_{in}-\sigma_{st}^{\bar{p}}$. The statistical error of the simulation is about $ 2\%$. With the exception of the first two points in Fig.8, corresponding to a momentum of 0.385 (0.3 – 0.47) and 0.56 (0.47 – 0.65) GeV/c, the experimental points in both figures fit well on the dashed curve corresponding to the cross section $\sigma_{in}-\sigma_{st}^{\bar{p}}$, when events are excluded with an antiproton which did not interact with the target nucleus. 

The measured inelastic cross-section is about 3 barns (Fig. 8 and Fig. 3 (c) in \cite{ALICE:2020zhb}  in the interval of the smallest momentum $0.30\le P\le 0.47$ GeV/c. It indicates a significant excess in comparison with the simulation results with the parameterization of the Glauber model used in GEANT4, by 2.1 times. It should be noted here that the quasi-classical approach underlying the Glauber and INC models demonstrates good agreement with experimental data for nuclear reactions at energies exceeding about 100 MeV/nucleon \cite{Barashenkov:1972zza}. In the low-energy range, other approaches to calculating cross-sections should be apply, such as the three-body model \cite{Paph:2005kei,Minomo:2016uto}.  It is also necessary to investigate the contribution of the Coulomb interaction, since, being a weakly coupled system, the antideutron is strongly polarized under the influence of the Coulomb field of the nucleus, which can lead to its breakup \cite{kharzeev1986kulonovskii2377}. Some contribution to the inelastic cross section of the antideutron-nucleus interaction at low energies can come from the antideutron diffraction splitting. It has been shown that in the energy range up to 200-300 MeV, the diffraction splitting of a deuteron makes a significant contribution to the cross section of the interaction with light nuclei \cite{Paph:2002kei}. For the antideutron this has yet to be explored.

\section{Some features of inelastic antideutron-nucleus interaction}

Experimental data on the antideutron-nucleus interaction are extremely scarce. To make sure that the model correctly describes this complex process, old data on the interaction of antideutrons at the momentum of 12.2 GeV/c with tantalum nuclei  were analyzed [4].  The data were obtained at the IHEP accelerator (Serpukhov). The antideutron beam had an admixture of $\pi^-$, which was about ${40\%}\pm{10\%}$. Events in which an antiproton spectator was identified among secondary charged particles were separated out because this process provides information on the antineutrons interaction  at the periphery of the target nucleus. But, in addition to the interaction of two antinucleons, the events with antineutron spectator have been also taken into consideration like antideutron-nucleus. 
 
\begin{table}
\centering
\begin{tabular}{|l|c|c|c|c|c|}
\hline
Reaction&$N_{tot}$&$<n_{ch}>$&$<n_{\pi^{-}}>$&$<n_{\pi_{ch}}>$&$<n_p>$\\
\hline
$(\bar{d}+\pi^{-})Ta$&1390&10.78&3.08&5.81&4.97 \\
experiment & &$\pm{0.15}$&$\pm{0.04}$&$\pm{0.07}$&$\pm{0.12}$ \\
\hline
$\bar{d}Ta$ tot &1500&$15.50$&$3.36$&
$6.15$&$8.99$ \\
\hline
$\bar{d}Ta$ no ${\bar{p}_{sp.}}$ &1171&$17.08$&$3.78$&
$6.84$&$10.08$ \\
\hline
$\bar{d}Ta$ cuts&1097&$10.74$&$3.29$&
$5.91$&$4.81$ \\
\hline
$\pi^{-} Ta$&1000&$9.02$&$1.77$&
$2.95$&$6.07$ \\
\hline
$\pi^{-} Ta$ cuts&820&$6.91$&$2.11$&
$3.54$&$3.37$ \\
\hline
$(\bar{d}+40\%\pi^{-})Ta$&&$9.65$&$2.95$&
$5.23$&$4.40$ \\
\hline
$(\bar{d}+30\%\pi^{-})Ta$&&$9.86$&$3.02$&
$5.36$&$4.48$ \\
\hline
\end{tabular}
\caption{\label{tab:widgets} Average multiplicities of secondary charged particles and calculation with INC for the $(\bar{d}+\pi^{-})Ta$ interactions at 12.2 GeV/c momentum. See the details in the text.}
\end{table}

The upper limiting momentum for all particles was chosen to be about 1 GeV/c. The lower limit of  secondary protons detection was determined by tantalum plate thickness and was about 300 MeV/c. Events were chosen with the multiplicity of secondary charged particles $n_{ch}\ge2$. Experimental data are presented in the line 1 of the Table 3. Results of simulation for inelastic  $\bar{d}Ta$ interaction without any selection and cuts are presented in the line 2. In the line 3 the events with antiproton spectator are excluded from consideration, but no any experimental cuts were applied. It is seen, that  antiproton stripping channel is about 22\%  of all inelastic interactions. In the line 4 there are  results of simulation, but with applying of all experimental  cut  and selections. In the lines 5 of the Table 3 there are  results  for $\pi^{-} Ta$ inelastic interaction without  experimental selections and cuts and in the line 6 are the same, but with experimental selections.   The results of calculations  with 40\%  and 30\%  admixture of $\pi^{-}Ta$ interaction with experimental selections and cuts are presented in the last two lines of the Table 3. Despite the high uncertainty of the experimental data associated with the admixture of $\pi^{-}Ta$  and the thickness of the tantalum plate, it can be concluded that the model as a whole describes this complex process quite satisfactory and can be used for some predictions for other energies of antideutron and target nuclei.

From simulation results it follows that stripping events with the presence of an antiproton-spectator  contribute  about 22\%. For an antineutron it should be about the same. That is, even for such a heavy nucleus like Ta, only in a little more than in half of the events (about 56\%) 2 antinucleons interact with the nucleus, and a little less than the half of the events (about 44\%) are the peripheral interaction of one antinucleon of the antideutron. 

Now let us consider, like an example,  the basic properties of  inelastic interaction of antideutrons with light and heavy nuclei of ALICE detector at the momentum 2 GeV/c. A large uncertainty in the interpretation of the experimental data is associated with the averaging for $Z$ and $A$ of the detector materials. For simulation the nuclei are taken $^{16}O$ close to average light nucleus $<Z>$=8.5, $<A>$=17.5  and $^{31}P$ close to  average  $<Z>=14.8$ and $<A>=31.8$. The results of ${10}^3$ events simulation are presented in the Table 4. The table shows the average multiplicities of secondary particles in three columns for each nucleus. The first column contains the values averaged over all inelastic interactions (100\% of all events).The second column shows the values for antiproton stripping events in which an antiproton spectator presents. The third column contains events without an antiproton spectator, that is, events with interaction of both antinucleons  and  peripheral antiproton interaction with antineutron spectator.

\begin{table}
\centering
\begin{tabular}{|l|c|c|c|c|c|c|}
\hline
Reaction&$^{16}O$&$^{16}O_{\bar{p}st}$&$^{16}O_{no\bar{p}st}$ &$^{31}P$&$^{31}P_{st}$&$^{31}P_{no\bar{p}st}$\\
\hline
$N,\%$&$100$&$36.02$&$63.98$&$100$&$32.62$&$67.38$\\
\hline
$<N_{ch}>$&6.09&$5.75$&$6.28$&$7.37$&$6.33$&$7.88$\\
\hline
$<N_{\pi_{ch}}>$&2.99&$2.38$&$3.34$&$3.15$&$2.25$&$3.58$\\
\hline
$<N_{\pi^-}>$&1.49&$1.03$&$1.76$&$1.59$&$0.99$&$1.88$\\
\hline
$<N_{p}>$&2.64&$2.35$&$2.80$&$3.82$&$3.07$&$4.18$\\
\hline
$<N_{\bar{p}}>$&$0.46$&$1.02$&$0.15$&$0.41$&$1.01$&$0.11$\\
\hline
$<N_{\bar{n}}>$&$0.46$&$0.12$&$0.65$&$0.40$&$0.13$&$0.53$\\
\hline
\end{tabular}
\caption{\label{tab:widgets} Average multiplicities of secondary charged particles calculated for the $\bar{d}+^{16}O$ and $\bar{d}+^{31}P$ interactions at 2 GeV/c momentum. See the details in the text.}
\end{table}

 First of all, it should be noted that for the lighter $^{16}O$ nucleus, about 36\% are events with the spectator antiproton.  For a heavier nucleus $^{31}P$, the number of such events decreases to about 32.6\%. The number of events with the  antineutron-spectator and peripheral antiproton  interaction should be approximately the same. This means that only in about 28\% of events for $^{16}O$ and 34.8\%  for $^{31}P$, 2 antinucleons interact with the nucleus.   It should be noted that the average multiplicity of antiprotons is slightly above one (1.01-1.02) in the events with an antiproton spectator. This is due to the presence of a charge exchange channel in the antineutron peripheral interaction. At the same time, the average yield of antineutrons indicates that approximately in 12\%  of inelastic antineutron interactions there is no  annihilation. The average total yield of antinucleons in the antideutron inelastic interaction  depends on the size of the target nucleus and is about 0.92 for $^{16}O$ and 0.81 for $^{31}P$. 

The figure 10 shows the distribution of all inelastic $\bar{d}+^{16}O$ interaction events  on the number of charged pions (thick solid histogram), which is the sum of the peripheral interaction of the antineutrons with  a spectator antiproton (dashed histogram) and summary interaction of ${\bar{p}}$  with ${\bar{n}}$-spectator and interaction of two antinucleons  (doted histogram).  The maximum of this distributions in the region of 2-4 charged pions mainly corresponds to the annihilation of one  antinucleon on the target nucleus. Assuming that the number of events with a antineutron-spectator is about the same as with an antiproton-spectator, the contribution of the processes with the interaction of two antinucleons can be approximately defined as the difference between the dotted and dashed histograms and presented on the figures  with thin solid histogram. It is shifted to higher multiplicity of charged pions and  has maximum at 4-6 and the tail up to 10 charged pions what corresponds to annihilation of two antinucleons. Our simulations show that the number  of events, when annihilation of two antinucleons occurred, is about 20\%  of all inelastic antideutron interactions with $^{16}O$ nucleus.

\begin{figure}
\centering
\includegraphics[width=0.950\linewidth]{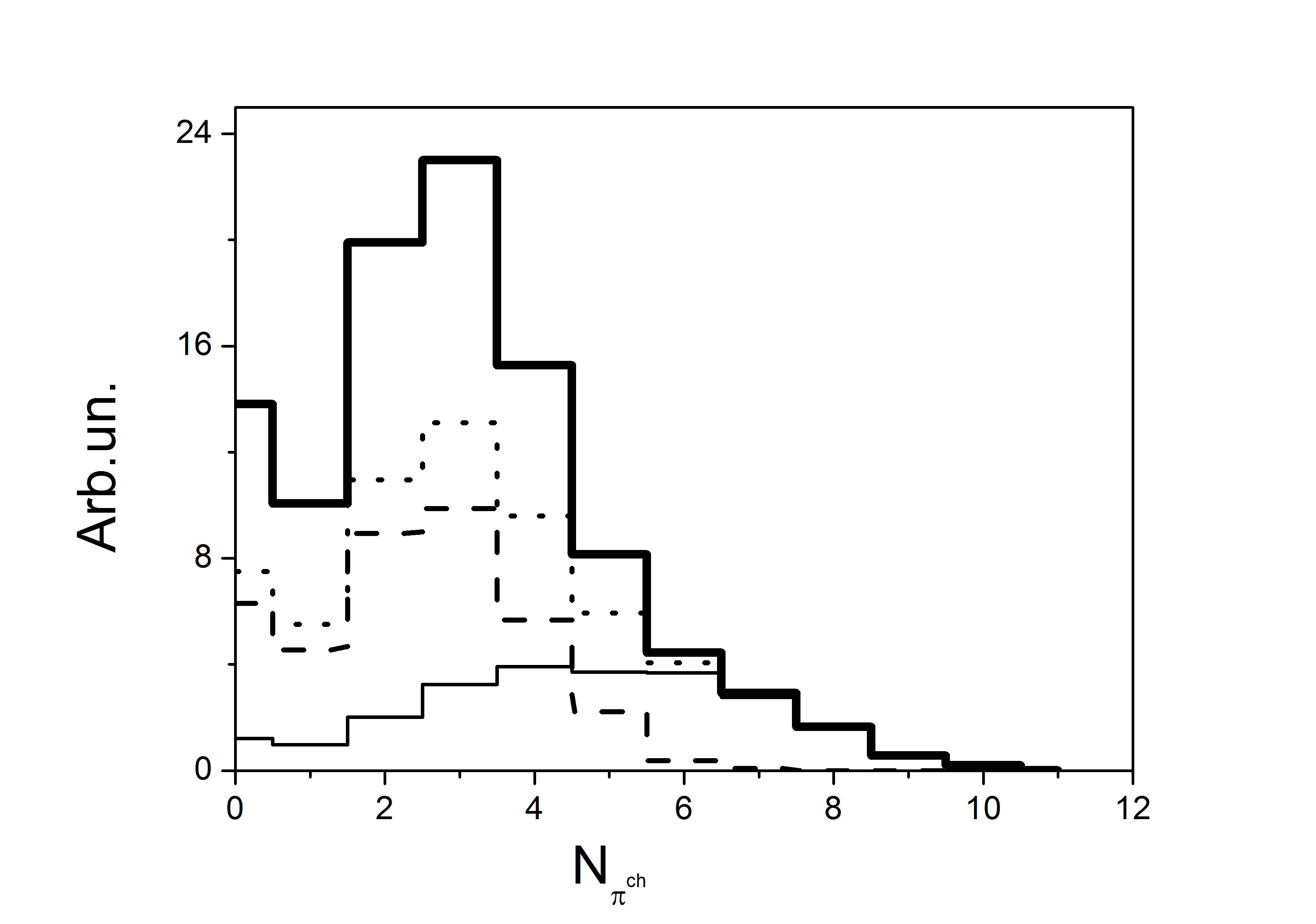}
\caption{Distribution on the number of charged pions  in the inelastic interaction $\bar{d}+^{16}O$ (thick solid histogram). The dashed histogram - stripping interaction with proton-spectator; the doted histogram - interaction of two antinucleons and interaction of ${\bar{p}}$  with ${\bar{n}}$-spectator, thin solid histogram - interaction of two antinucleons.}\label{fig_10.jpg}
\end{figure}

\begin{figure}
\centering
\includegraphics[width=0.950\linewidth]{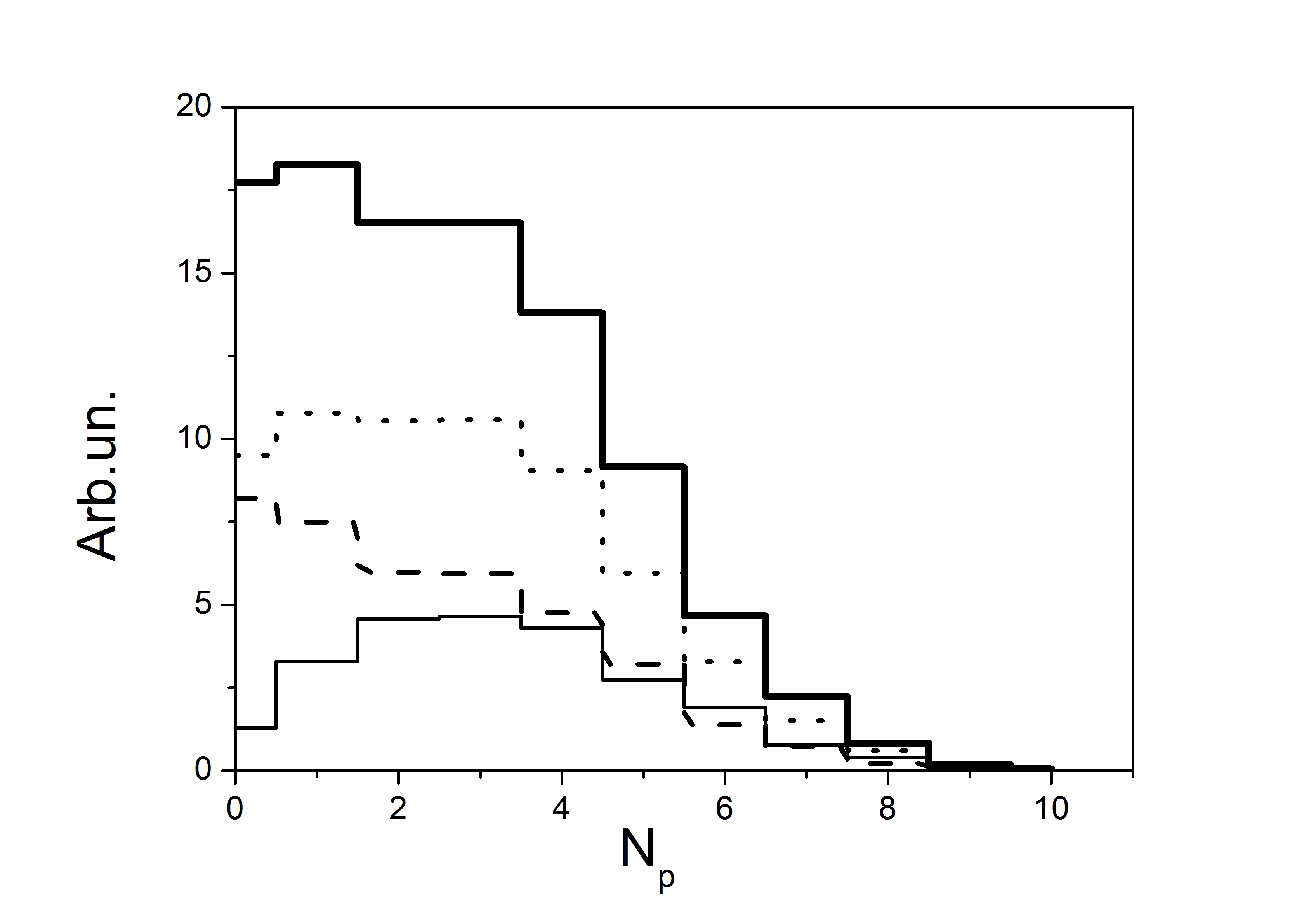}
\caption{Distribution on the number of protons in $\bar{d}+^{16}O$ inelastic interaction. Notation as shown in Fig. 10.}\label{fig_11.jpg}
\end{figure}

\begin{figure}
\centering
\includegraphics[width=0.950\linewidth]{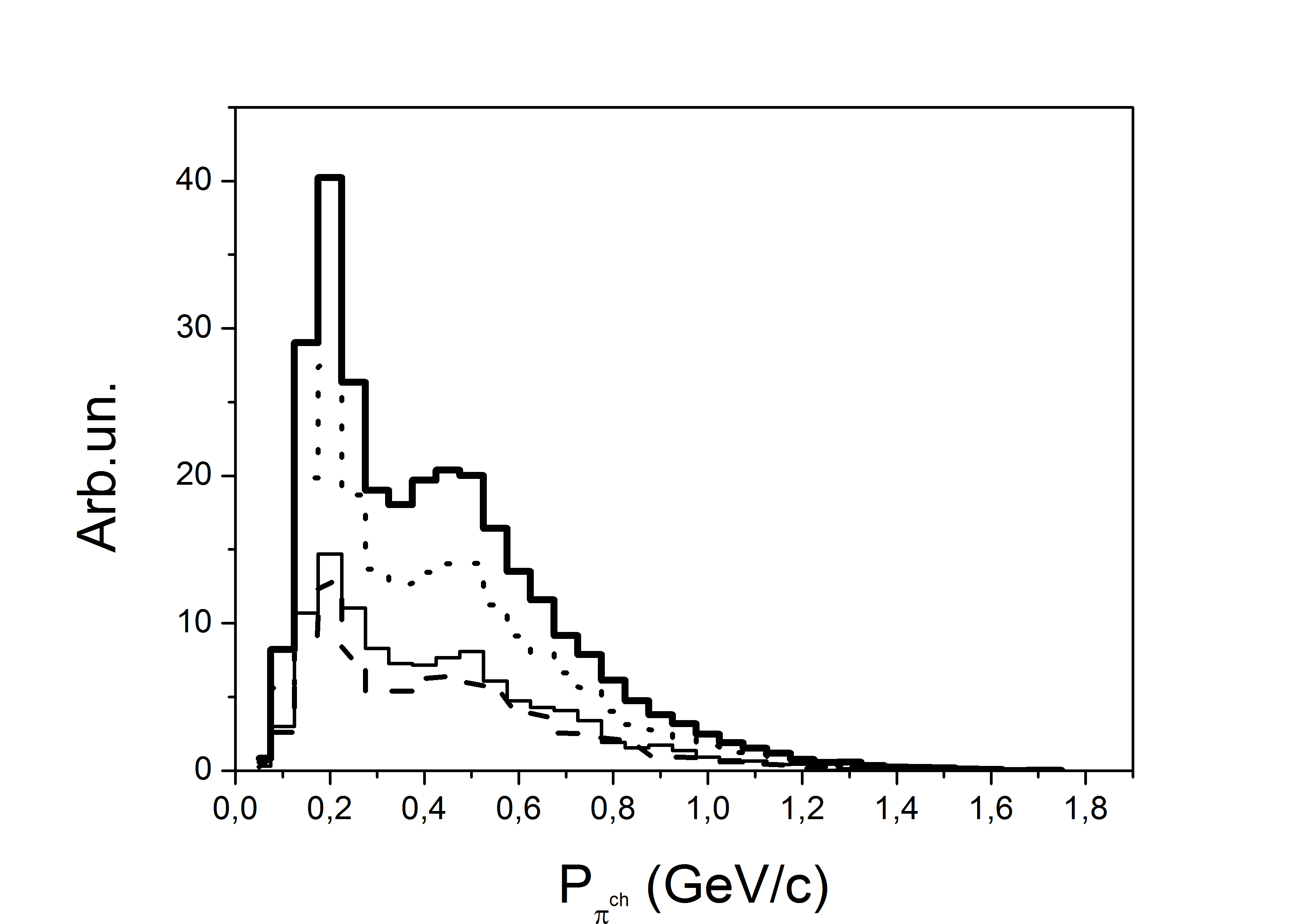}
\caption{Momentum distribution of charged pions in $\bar{d}+^{16}O$ inelastic interaction. Notation as shown in Fig. 10.}\label{fig_12.jpg}
\end{figure}

\begin{figure}
\centering
\includegraphics[width=0.950\linewidth]{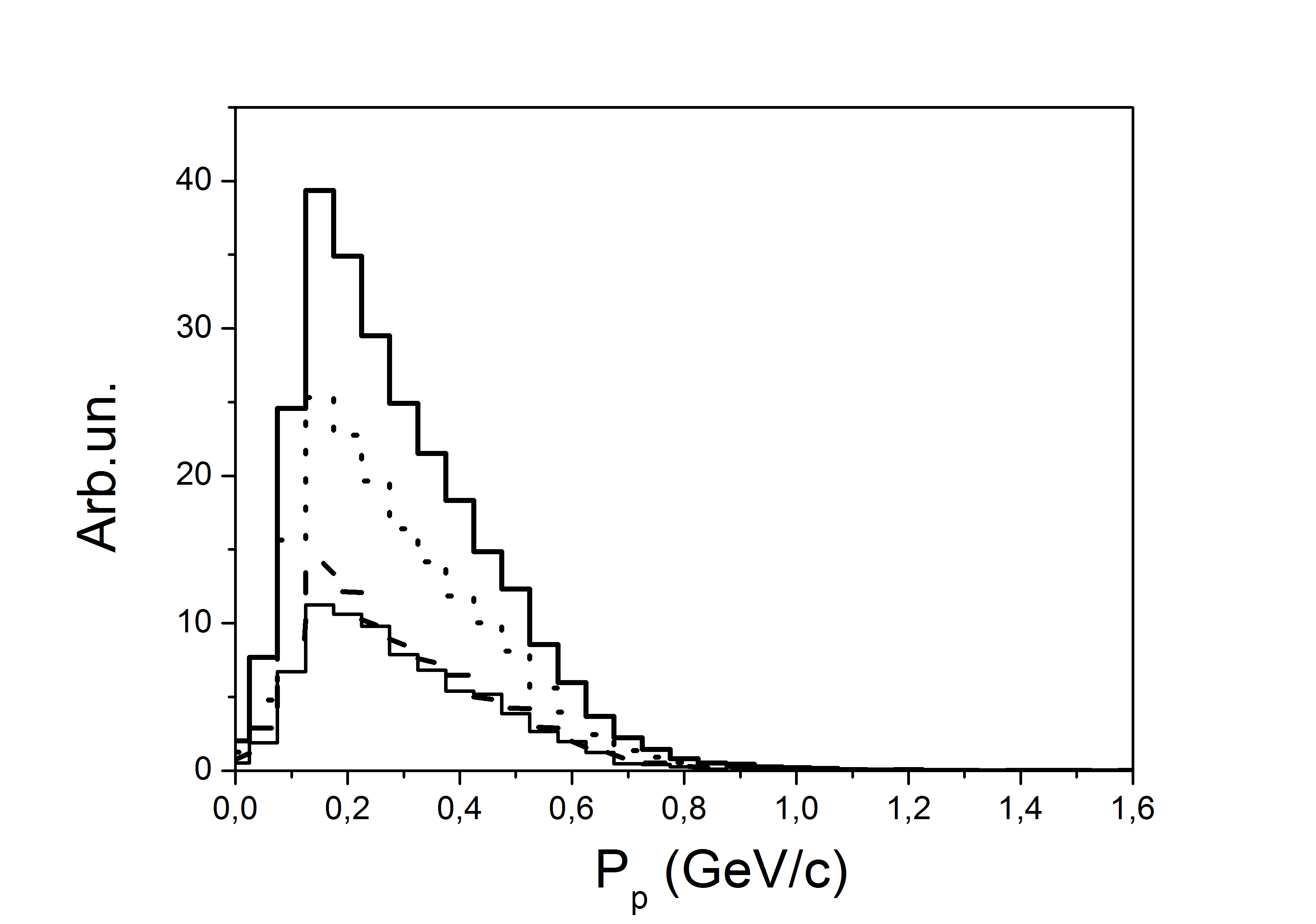}
\caption{Momentum distribution of protons in $\bar{d}+^{16}O$ inelastic interaction. Notation as shown in Fig. 10.}\label{fig_13.jpg}
\end{figure}

The figure 11 shows the distribution of all inelastic $\bar{d}+^{16}O$ interaction events  on the number of protons. As expected, the number of events with a large number of ejected protons  is noticeably lower in events with peripheral interaction of an antineutron and spectator antiproton (dashed histogram). In the case of annihilation of two antinucleons (thin solid histogram) the maximum of the distribution is at the values of 2 - 3 protons. In addition, it can be seen that there is a number of events when more than 8 protons originate from the oxygen nucleus. This is possible as a result of the charge exchange  of  $\pi^+$ and $\pi^0$ on the neutron of the target nucleus $(\pi^{+,0}+n)\to(\pi^{0,-}+p)$. In the case of annihilation of both antinucleons a complete destroy of such a light nucleus as $^{16}O$  is possible.

Momentum distributions of charged pions are demonstrated on Fig.12. The  part  of the distribution with  higher momentum with maximum around  0.5 GeV/c corresponds to the momentum of pions in annihilation on intranuclear nucleons. The maximum of the distribution in the region of 0.2 GeV/c corresponds to the momentum of the pions after the transfer the energy to the nucleus as a result of  FSI. This form of momentum distribution is typical for all types of events.

Figure 13 shows the momentum distribution of  ejected protons. The high-momentum part of this distribution corresponds to protons knocked out from the nucleus by annihilation pions and extends up to 0.9-1 GeV/c with both  the peripheral interaction of an antineutron and in the annihilation of two antinucleons. The low-momentum part of the distribution is determined mainly by evaporation protons in the processes of de-excitation of residual nucleus after the fast cascade stage.

The presented simulation results for inelastic interaction of antideutrons with nuclei  clearly demonstrate that the peripheral interaction of one antinucleon  with the nucleus is a significant channel of the inelastic interaction of antideutron (about 72\% of all events for $^{16}O$ and 65\%  for -  $^{31}P$). Annihilation of both  antinucleons occurs only in about 20\%  of all events for $^{16}O$ and in 29\%  for $^{31}P$ at the antideutron momentum 2 GeV/c.  Similar simulations can be performed for antideutrons in the energy range $100$  MeV/n $\le T_{kin}\le 25-30$ GeV/n and for nuclei starting from $^{12}C$.  It is very important that the proposed model of antideutron-nucleus interaction can be improved based on experimental data for deutron-nucleus interaction.

\section{Conclution}

 The new version of INC model  proposed for simulation the  antideutron-nucleus inelastic interaction in the region of antideutron energies of $100$  MeV/n $\le T_{kin}\le 25-30$ GeV/n for a wide range of target nuclei. Even at the first steps the model demonstrates a good agreement with experimental data for cross sections of inelastic interactions of (anti)protons and deuterons with nuclei. Without any additional parameters, the model is in satisfactory agreement with the available experimental data for inelastic antideutron-nucleus interaction. The cross sections of inelastic interaction obtained within the framework of the proposed model can be used for simulation the propagation of antideutrons through galactic matter from their source to the detection point near the Earth. Cross-sections of light antinuclei inelastic interaction are important for the development of cosmic ray physics\cite{maurin2025precisioncrosssectionsadvancingcosmicray}, but measuring of antinuclei cross-sections is a complex and lengthy experimental task. The suggested model can be further tested and improved provided availability of more data for (anti)nucleon-nucleus and deutron-nucleus interaction.

  Some devices, such as AMS \cite{Kounine:2012ega} and GAPS \cite{Hailey:2009fpa}, are designed specially to detect light antinuclei. Detailed simulations of light antinuclei interaction with nuclei are important  for the experiments on their detection. For such experiments, information is needed on the multiplicity and spectra of secondary particles after the interaction of light antinuclei with detector. Based on the proposed model, Monte Carlo (MC) generator has been created for the simulation of the inelastic interaction of antideutrons with nuclei. This generator can be applied in the preparation and analysis of experiments to search for antideutrons near the Earth.  

\bibliography{bibliography}

\begin{thebibliography}{38}%
\makeatletter
\providecommand \@ifxundefined [1]{%
 \@ifx{#1\undefined}
}%
\providecommand \@ifnum [1]{%
 \ifnum #1\expandafter \@firstoftwo
 \else \expandafter \@secondoftwo
 \fi
}%
\providecommand \@ifx [1]{%
 \ifx #1\expandafter \@firstoftwo
 \else \expandafter \@secondoftwo
 \fi
}%
\providecommand \natexlab [1]{#1}%
\providecommand \enquote  [1]{``#1''}%
\providecommand \bibnamefont  [1]{#1}%
\providecommand \bibfnamefont [1]{#1}%
\providecommand \citenamefont [1]{#1}%
\providecommand \href@noop [0]{\@secondoftwo}%
\providecommand \href [0]{\begingroup \@sanitize@url \@href}%
\providecommand \@href[1]{\@@startlink{#1}\@@href}%
\providecommand \@@href[1]{\endgroup#1\@@endlink}%
\providecommand \@sanitize@url [0]{\catcode `\\12\catcode `\$12\catcode `\&12\catcode `\#12\catcode `\^12\catcode `\_12\catcode `\%12\relax}%
\providecommand \@@startlink[1]{}%
\providecommand \@@endlink[0]{}%
\providecommand \url  [0]{\begingroup\@sanitize@url \@url }%
\providecommand \@url [1]{\endgroup\@href {#1}{\urlprefix }}%
\providecommand \urlprefix  [0]{URL }%
\providecommand \Eprint [0]{\href }%
\providecommand \doibase [0]{https://doi.org/}%
\providecommand \selectlanguage [0]{\@gobble}%
\providecommand \bibinfo  [0]{\@secondoftwo}%
\providecommand \bibfield  [0]{\@secondoftwo}%
\providecommand \translation [1]{[#1]}%
\providecommand \BibitemOpen [0]{}%
\providecommand \bibitemStop [0]{}%
\providecommand \bibitemNoStop [0]{.\EOS\space}%
\providecommand \EOS [0]{\spacefactor3000\relax}%
\providecommand \BibitemShut  [1]{\csname bibitem#1\endcsname}%
\let\auto@bib@innerbib\@empty
\bibitem [{\citenamefont {Richard}(2023)}]{Richard:2022tpn}%
  \BibitemOpen
  \bibfield  {author} {\bibinfo {author} {\bibfnamefont {J.-M.}\ \bibnamefont {Richard}},\ }\bibinfo {title} {{Nucleon-Antinucleon Interaction}},\ in\ \href {https://doi.org/10.1007/978-981-19-6345-2_53} {\emph {\bibinfo {booktitle} {{Handbook of Nuclear Physics}}}},\ \bibinfo {editor} {edited by\ \bibinfo {editor} {\bibfnamefont {I.}~\bibnamefont {Tanihata}}, \bibinfo {editor} {\bibfnamefont {H.}~\bibnamefont {Toki}},\ and\ \bibinfo {editor} {\bibfnamefont {T.}~\bibnamefont {Kajino}}}\ (\bibinfo {year} {2023})\ pp.\ \bibinfo {pages} {1--22},\ \Eprint {https://arxiv.org/abs/2205.02529} {arXiv:2205.02529 [nucl-th]} \BibitemShut {NoStop}%
\bibitem [{\citenamefont {Denisov}\ \emph {et~al.}(1971)\citenamefont {Denisov}, \citenamefont {Donskov}, \citenamefont {Gorin}, \citenamefont {Kachanov}, \citenamefont {Kutjin}, \citenamefont {Petrukhin}, \citenamefont {Prokoshkin}, \citenamefont {Razuvaev}, \citenamefont {Shuvalov},\ and\ \citenamefont {Stojanova}}]{Denisov:1971im}%
  \BibitemOpen
  \bibfield  {author} {\bibinfo {author} {\bibfnamefont {S.~P.}\ \bibnamefont {Denisov}}, \bibinfo {author} {\bibfnamefont {S.~V.}\ \bibnamefont {Donskov}}, \bibinfo {author} {\bibfnamefont {Y.~P.}\ \bibnamefont {Gorin}}, \bibinfo {author} {\bibfnamefont {V.~A.}\ \bibnamefont {Kachanov}}, \bibinfo {author} {\bibfnamefont {V.~M.}\ \bibnamefont {Kutjin}}, \bibinfo {author} {\bibfnamefont {A.~I.}\ \bibnamefont {Petrukhin}}, \bibinfo {author} {\bibfnamefont {Y.~D.}\ \bibnamefont {Prokoshkin}}, \bibinfo {author} {\bibfnamefont {E.~A.}\ \bibnamefont {Razuvaev}}, \bibinfo {author} {\bibfnamefont {R.~S.}\ \bibnamefont {Shuvalov}},\ and\ \bibinfo {author} {\bibfnamefont {D.~A.}\ \bibnamefont {Stojanova}},\ }\bibfield  {title} {\bibinfo {title} {{Measurements of anti-deuteron absorption and stripping cross sections at the momentum 13.3 gev/c}},\ }\href {https://doi.org/10.1016/0550-3213(71)90229-X} {\bibfield  {journal} {\bibinfo  {journal} {Nucl. Phys. B}\ }\textbf {\bibinfo {volume} {31}},\ \bibinfo {pages} {253}
  (\bibinfo {year} {1971})}\BibitemShut {NoStop}%
\bibitem [{\citenamefont {Binon}\ \emph {et~al.}(1970)\citenamefont {Binon} \emph {et~al.}}]{Binon:1970yu}%
  \BibitemOpen
  \bibfield  {author} {\bibinfo {author} {\bibfnamefont {F.~G.}\ \bibnamefont {Binon}} \emph {et~al.},\ }\bibfield  {title} {\bibinfo {title} {{Absorption cross-sections of 25 gev/c antideuterons in li, c, al, cu and pb}},\ }\href {https://doi.org/10.1016/0370-2693(70)90112-7} {\bibfield  {journal} {\bibinfo  {journal} {Phys. Lett. B}\ }\textbf {\bibinfo {volume} {31}},\ \bibinfo {pages} {230} (\bibinfo {year} {1970})}\BibitemShut {NoStop}%
\bibitem [{\citenamefont {Andreev}\ \emph {et~al.}(1990)\citenamefont {Andreev}, \citenamefont {Baranov}, \citenamefont {Golubeva}, \citenamefont {Ilinov}, \citenamefont {Levonian}, \citenamefont {Pol},\ and\ \citenamefont {Filkov}}]{Andreev:1990mk}%
  \BibitemOpen
  \bibfield  {author} {\bibinfo {author} {\bibfnamefont {V.~F.}\ \bibnamefont {Andreev}}, \bibinfo {author} {\bibfnamefont {P.~S.}\ \bibnamefont {Baranov}}, \bibinfo {author} {\bibfnamefont {E.~S.}\ \bibnamefont {Golubeva}}, \bibinfo {author} {\bibfnamefont {A.~S.}\ \bibnamefont {Ilinov}}, \bibinfo {author} {\bibfnamefont {S.~V.}\ \bibnamefont {Levonian}}, \bibinfo {author} {\bibfnamefont {Y.~S.}\ \bibnamefont {Pol}},\ and\ \bibinfo {author} {\bibfnamefont {L.~V.}\ \bibnamefont {Filkov}},\ }\bibfield  {title} {\bibinfo {title} {{Multiplicities and correlations of secondary charged particles in the interactions of anti-neutrons and anti-deuterons with a momentum of 6.1-GeV/c per nucleon with tantalum nuclei. (In Russian)}},\ }\href@noop {} {\bibfield  {journal} {\bibinfo  {journal} {Sov. J. Nucl. Phys.}\ }\textbf {\bibinfo {volume} {51}},\ \bibinfo {pages} {88} (\bibinfo {year} {1990})}\BibitemShut {NoStop}%
\bibitem [{\citenamefont {Armstrong}\ \emph {et~al.}(2000)\citenamefont {Armstrong} \emph {et~al.}}]{E864:2000loc}%
  \BibitemOpen
  \bibfield  {author} {\bibinfo {author} {\bibfnamefont {T.~A.}\ \bibnamefont {Armstrong}} \emph {et~al.} (\bibinfo {collaboration} {E864}),\ }\bibfield  {title} {\bibinfo {title} {{Anti-deuteron yield at the AGS and coalescence implications}},\ }\href {https://doi.org/10.1103/PhysRevLett.85.2685} {\bibfield  {journal} {\bibinfo  {journal} {Phys. Rev. Lett.}\ }\textbf {\bibinfo {volume} {85}},\ \bibinfo {pages} {2685} (\bibinfo {year} {2000})},\ \Eprint {https://arxiv.org/abs/nucl-ex/0005001} {arXiv:nucl-ex/0005001} \BibitemShut {NoStop}%
\bibitem [{\citenamefont {Adler}\ \emph {et~al.}(2001)\citenamefont {Adler} \emph {et~al.}}]{STAR:2001pbk}%
  \BibitemOpen
  \bibfield  {author} {\bibinfo {author} {\bibfnamefont {C.}~\bibnamefont {Adler}} \emph {et~al.} (\bibinfo {collaboration} {STAR}),\ }\bibfield  {title} {\bibinfo {title} {{Anti-deuteron and anti-He-3 production in s(NN)**(1/2) = 130-GeV Au+Au collisions}},\ }\href {https://doi.org/10.1103/PhysRevLett.87.262301} {\bibfield  {journal} {\bibinfo  {journal} {Phys. Rev. Lett.}\ }\textbf {\bibinfo {volume} {87}},\ \bibinfo {pages} {262301} (\bibinfo {year} {2001})},\ \bibinfo {note} {[Erratum: Phys.Rev.Lett. 87, 279902 (2001)]},\ \Eprint {https://arxiv.org/abs/nucl-ex/0108022} {arXiv:nucl-ex/0108022} \BibitemShut {NoStop}%
\bibitem [{\citenamefont {Adler}\ \emph {et~al.}(2005)\citenamefont {Adler} \emph {et~al.}}]{PHENIX:2004vqi}%
  \BibitemOpen
  \bibfield  {author} {\bibinfo {author} {\bibfnamefont {S.~S.}\ \bibnamefont {Adler}} \emph {et~al.} (\bibinfo {collaboration} {PHENIX}),\ }\bibfield  {title} {\bibinfo {title} {{Deuteron and antideuteron production in Au + Au collisions at s(NN)**(1/2) = 200-GeV}},\ }\href {https://doi.org/10.1103/PhysRevLett.94.122302} {\bibfield  {journal} {\bibinfo  {journal} {Phys. Rev. Lett.}\ }\textbf {\bibinfo {volume} {94}},\ \bibinfo {pages} {122302} (\bibinfo {year} {2005})},\ \Eprint {https://arxiv.org/abs/nucl-ex/0406004} {arXiv:nucl-ex/0406004} \BibitemShut {NoStop}%
\bibitem [{\citenamefont {Anticic}\ \emph {et~al.}(2012)\citenamefont {Anticic} \emph {et~al.}}]{NA49:2011blr}%
  \BibitemOpen
  \bibfield  {author} {\bibinfo {author} {\bibfnamefont {T.}~\bibnamefont {Anticic}} \emph {et~al.} (\bibinfo {collaboration} {NA49}),\ }\bibfield  {title} {\bibinfo {title} {{Antideuteron and deuteron production in mid-central Pb+Pb collisions at 158$A$ GeV}},\ }\href {https://doi.org/10.1103/PhysRevC.85.044913} {\bibfield  {journal} {\bibinfo  {journal} {Phys. Rev. C}\ }\textbf {\bibinfo {volume} {85}},\ \bibinfo {pages} {044913} (\bibinfo {year} {2012})},\ \Eprint {https://arxiv.org/abs/1111.2588} {arXiv:1111.2588 [nucl-ex]} \BibitemShut {NoStop}%
\bibitem [{\citenamefont {Adam}\ \emph {et~al.}(2019)\citenamefont {Adam} \emph {et~al.}}]{STAR:2019sjh}%
  \BibitemOpen
  \bibfield  {author} {\bibinfo {author} {\bibfnamefont {J.}~\bibnamefont {Adam}} \emph {et~al.} (\bibinfo {collaboration} {STAR}),\ }\bibfield  {title} {\bibinfo {title} {{Beam energy dependence of (anti-)deuteron production in Au + Au collisions at the BNL Relativistic Heavy Ion Collider}},\ }\href {https://doi.org/10.1103/PhysRevC.99.064905} {\bibfield  {journal} {\bibinfo  {journal} {Phys. Rev. C}\ }\textbf {\bibinfo {volume} {99}},\ \bibinfo {pages} {064905} (\bibinfo {year} {2019})},\ \Eprint {https://arxiv.org/abs/1903.11778} {arXiv:1903.11778 [nucl-ex]} \BibitemShut {NoStop}%
\bibitem [{\citenamefont {Acharya}\ \emph {et~al.}(2017)\citenamefont {Acharya} \emph {et~al.}}]{ALICE:2017nuf}%
  \BibitemOpen
  \bibfield  {author} {\bibinfo {author} {\bibfnamefont {S.}~\bibnamefont {Acharya}} \emph {et~al.} (\bibinfo {collaboration} {ALICE}),\ }\bibfield  {title} {\bibinfo {title} {{Measurement of deuteron spectra and elliptic flow in Pb\textendash{}Pb collisions at $\sqrt{s_{\mathrm {NN}}}$ = 2.76 TeV at the LHC}},\ }\href {https://doi.org/10.1140/epjc/s10052-017-5222-x} {\bibfield  {journal} {\bibinfo  {journal} {Eur. Phys. J. C}\ }\textbf {\bibinfo {volume} {77}},\ \bibinfo {pages} {658} (\bibinfo {year} {2017})},\ \Eprint {https://arxiv.org/abs/1707.07304} {arXiv:1707.07304 [nucl-ex]} \BibitemShut {NoStop}%
\bibitem [{\citenamefont {Acharya}\ \emph {et~al.}(2020{\natexlab{a}})\citenamefont {Acharya} \emph {et~al.}}]{ALICE:2019ikx}%
  \BibitemOpen
  \bibfield  {author} {\bibinfo {author} {\bibfnamefont {S.}~\bibnamefont {Acharya}} \emph {et~al.} (\bibinfo {collaboration} {ALICE}),\ }\bibfield  {title} {\bibinfo {title} {{Measurement of the (anti-)3He elliptic flow in Pb\textendash{}Pb collisions at sNN=5.02TeV}},\ }\href {https://doi.org/10.1016/j.physletb.2020.135414} {\bibfield  {journal} {\bibinfo  {journal} {Phys. Lett. B}\ }\textbf {\bibinfo {volume} {805}},\ \bibinfo {pages} {135414} (\bibinfo {year} {2020}{\natexlab{a}})},\ \Eprint {https://arxiv.org/abs/1910.09718} {arXiv:1910.09718 [nucl-ex]} \BibitemShut {NoStop}%
\bibitem [{\citenamefont {Acharya}\ \emph {et~al.}(2020{\natexlab{b}})\citenamefont {Acharya} \emph {et~al.}}]{ALICE:2020chv}%
  \BibitemOpen
  \bibfield  {author} {\bibinfo {author} {\bibfnamefont {S.}~\bibnamefont {Acharya}} \emph {et~al.} (\bibinfo {collaboration} {ALICE}),\ }\bibfield  {title} {\bibinfo {title} {{Elliptic and triangular flow of (anti)deuterons in Pb-Pb collisions at $\sqrt{s_{\mathrm{NN}}}$ = 5.02 TeV}},\ }\href {https://doi.org/10.1103/PhysRevC.102.055203} {\bibfield  {journal} {\bibinfo  {journal} {Phys. Rev. C}\ }\textbf {\bibinfo {volume} {102}},\ \bibinfo {pages} {055203} (\bibinfo {year} {2020}{\natexlab{b}})},\ \Eprint {https://arxiv.org/abs/2005.14639} {arXiv:2005.14639 [nucl-ex]} \BibitemShut {NoStop}%
\bibitem [{\citenamefont {Acharya}\ \emph {et~al.}(2023)\citenamefont {Acharya} \emph {et~al.}}]{ALICE:2022xiu}%
  \BibitemOpen
  \bibfield  {author} {\bibinfo {author} {\bibfnamefont {S.}~\bibnamefont {Acharya}} \emph {et~al.} (\bibinfo {collaboration} {ALICE}),\ }\bibfield  {title} {\bibinfo {title} {{First Measurement of Antideuteron Number Fluctuations at Energies Available at the Large Hadron Collider}},\ }\href {https://doi.org/10.1103/PhysRevLett.131.041901} {\bibfield  {journal} {\bibinfo  {journal} {Phys. Rev. Lett.}\ }\textbf {\bibinfo {volume} {131}},\ \bibinfo {pages} {041901} (\bibinfo {year} {2023})},\ \Eprint {https://arxiv.org/abs/2204.10166} {arXiv:2204.10166 [nucl-ex]} \BibitemShut {NoStop}%
\bibitem [{\citenamefont {Acharya}\ \emph {et~al.}(2020{\natexlab{c}})\citenamefont {Acharya} \emph {et~al.}}]{ALICE:2020zhb}%
  \BibitemOpen
  \bibfield  {author} {\bibinfo {author} {\bibfnamefont {S.}~\bibnamefont {Acharya}} \emph {et~al.} (\bibinfo {collaboration} {ALICE}),\ }\bibfield  {title} {\bibinfo {title} {{Measurement of the low-energy antideuteron inelastic cross section}},\ }\href {https://doi.org/10.1103/PhysRevLett.125.162001} {\bibfield  {journal} {\bibinfo  {journal} {Phys. Rev. Lett.}\ }\textbf {\bibinfo {volume} {125}},\ \bibinfo {pages} {162001} (\bibinfo {year} {2020}{\natexlab{c}})},\ \Eprint {https://arxiv.org/abs/2005.11122} {arXiv:2005.11122 [nucl-ex]} \BibitemShut {NoStop}%
\bibitem [{\citenamefont {Aramaki}\ \emph {et~al.}(2016)\citenamefont {Aramaki} \emph {et~al.}}]{Aramaki:2015pii}%
  \BibitemOpen
  \bibfield  {author} {\bibinfo {author} {\bibfnamefont {T.}~\bibnamefont {Aramaki}} \emph {et~al.},\ }\bibfield  {title} {\bibinfo {title} {{Review of the theoretical and experimental status of dark matter identification with cosmic-ray antideuterons}},\ }\href {https://doi.org/10.1016/j.physrep.2016.01.002} {\bibfield  {journal} {\bibinfo  {journal} {Phys. Rept.}\ }\textbf {\bibinfo {volume} {618}},\ \bibinfo {pages} {1} (\bibinfo {year} {2016})},\ \Eprint {https://arxiv.org/abs/1505.07785} {arXiv:1505.07785 [hep-ph]} \BibitemShut {NoStop}%
\bibitem [{\citenamefont {Donato}\ \emph {et~al.}(2008)\citenamefont {Donato}, \citenamefont {Fornengo},\ and\ \citenamefont {Maurin}}]{Donato:2008yx}%
  \BibitemOpen
  \bibfield  {author} {\bibinfo {author} {\bibfnamefont {F.}~\bibnamefont {Donato}}, \bibinfo {author} {\bibfnamefont {N.}~\bibnamefont {Fornengo}},\ and\ \bibinfo {author} {\bibfnamefont {D.}~\bibnamefont {Maurin}},\ }\bibfield  {title} {\bibinfo {title} {{Antideuteron fluxes from dark matter annihilation in diffusion models}},\ }\href {https://doi.org/10.1103/PhysRevD.78.043506} {\bibfield  {journal} {\bibinfo  {journal} {Phys. Rev. D}\ }\textbf {\bibinfo {volume} {78}},\ \bibinfo {pages} {043506} (\bibinfo {year} {2008})},\ \Eprint {https://arxiv.org/abs/0803.2640} {arXiv:0803.2640 [hep-ph]} \BibitemShut {NoStop}%
\bibitem [{\citenamefont {Baer}\ and\ \citenamefont {Profumo}(2005)}]{Baer_2005}%
  \BibitemOpen
  \bibfield  {author} {\bibinfo {author} {\bibfnamefont {H.}~\bibnamefont {Baer}}\ and\ \bibinfo {author} {\bibfnamefont {S.}~\bibnamefont {Profumo}},\ }\bibfield  {title} {\bibinfo {title} {Low energy antideuterons: shedding light on dark matter},\ }\href {https://doi.org/10.1088/1475-7516/2005/12/008} {\bibfield  {journal} {\bibinfo  {journal} {Journal of Cosmology and Astroparticle Physics}\ }\textbf {\bibinfo {volume} {2005}}\bibinfo  {number} { (12)},\ \bibinfo {pages} {008–008}}\BibitemShut {NoStop}%
\bibitem [{\citenamefont {Silk}\ \emph {et~al.}(2010)\citenamefont {Silk} \emph {et~al.}}]{Bertone:2010zza}%
  \BibitemOpen
\bibfield  {number} {  }\bibfield  {author} {\bibinfo {author} {\bibfnamefont {J.}~\bibnamefont {Silk}} \emph {et~al.},\ }\href {https://doi.org/10.1017/CBO9780511770739} {\emph {\bibinfo {title} {{Particle Dark Matter: Observations, Models and Searches}}}},\ edited by\ \bibinfo {editor} {\bibfnamefont {G.}~\bibnamefont {Bertone}}\ (\bibinfo  {publisher} {Cambridge Univ. Press},\ \bibinfo {address} {Cambridge},\ \bibinfo {year} {2010})\BibitemShut {NoStop}%
\bibitem [{\citenamefont {Ibarra}\ and\ \citenamefont {Wild}(2013)}]{Ibarra:2012cc}%
  \BibitemOpen
  \bibfield  {author} {\bibinfo {author} {\bibfnamefont {A.}~\bibnamefont {Ibarra}}\ and\ \bibinfo {author} {\bibfnamefont {S.}~\bibnamefont {Wild}},\ }\bibfield  {title} {\bibinfo {title} {{Prospects of antideuteron detection from dark matter annihilations or decays at AMS-02 and GAPS}},\ }\href {https://doi.org/10.1088/1475-7516/2013/02/021} {\bibfield  {journal} {\bibinfo  {journal} {JCAP}\ }\textbf {\bibinfo {volume} {02}},\ \bibinfo {pages} {021}},\ \Eprint {https://arxiv.org/abs/1209.5539} {arXiv:1209.5539 [hep-ph]} \BibitemShut {NoStop}%
\bibitem [{\citenamefont {Donato}\ \emph {et~al.}(2000)\citenamefont {Donato}, \citenamefont {Fornengo},\ and\ \citenamefont {Salati}}]{Donato:1999gy}%
  \BibitemOpen
  \bibfield  {author} {\bibinfo {author} {\bibfnamefont {F.}~\bibnamefont {Donato}}, \bibinfo {author} {\bibfnamefont {N.}~\bibnamefont {Fornengo}},\ and\ \bibinfo {author} {\bibfnamefont {P.}~\bibnamefont {Salati}},\ }\bibfield  {title} {\bibinfo {title} {{Anti-deuterons as a signature of supersymmetric dark matter}},\ }\href {https://doi.org/10.1103/PhysRevD.62.043003} {\bibfield  {journal} {\bibinfo  {journal} {Phys. Rev. D}\ }\textbf {\bibinfo {volume} {62}},\ \bibinfo {pages} {043003} (\bibinfo {year} {2000})},\ \Eprint {https://arxiv.org/abs/hep-ph/9904481} {arXiv:hep-ph/9904481} \BibitemShut {NoStop}%
\bibitem [{\citenamefont {Kounine}(2012)}]{Kounine:2012ega}%
  \BibitemOpen
  \bibfield  {author} {\bibinfo {author} {\bibfnamefont {A.}~\bibnamefont {Kounine}},\ }\bibfield  {title} {\bibinfo {title} {{The Alpha Magnetic Spectrometer on the International Space Station}},\ }\href {https://doi.org/10.1142/S0218301312300056} {\bibfield  {journal} {\bibinfo  {journal} {Int. J. Mod. Phys. E}\ }\textbf {\bibinfo {volume} {21}},\ \bibinfo {pages} {1230005} (\bibinfo {year} {2012})}\BibitemShut {NoStop}%
\bibitem [{\citenamefont {Hailey}(2009)}]{Hailey:2009fpa}%
  \BibitemOpen
  \bibfield  {author} {\bibinfo {author} {\bibfnamefont {C.~J.}\ \bibnamefont {Hailey}},\ }\bibfield  {title} {\bibinfo {title} {{An indirect search for dark matter using antideuterons: the GAPS experiment}},\ }\href {https://doi.org/10.1088/1367-2630/11/10/105022} {\bibfield  {journal} {\bibinfo  {journal} {New J. Phys.}\ }\textbf {\bibinfo {volume} {11}},\ \bibinfo {pages} {105022} (\bibinfo {year} {2009})}\BibitemShut {NoStop}%
\bibitem [{\citenamefont {Fornengo}\ \emph {et~al.}(2013)\citenamefont {Fornengo}, \citenamefont {Maccione},\ and\ \citenamefont {Vittino}}]{Fornengo:2013osa}%
  \BibitemOpen
  \bibfield  {author} {\bibinfo {author} {\bibfnamefont {N.}~\bibnamefont {Fornengo}}, \bibinfo {author} {\bibfnamefont {L.}~\bibnamefont {Maccione}},\ and\ \bibinfo {author} {\bibfnamefont {A.}~\bibnamefont {Vittino}},\ }\bibfield  {title} {\bibinfo {title} {{Dark matter searches with cosmic antideuterons: status and perspectives}},\ }\href {https://doi.org/10.1088/1475-7516/2013/09/031} {\bibfield  {journal} {\bibinfo  {journal} {JCAP}\ }\textbf {\bibinfo {volume} {09}},\ \bibinfo {pages} {031}},\ \Eprint {https://arxiv.org/abs/1306.4171} {arXiv:1306.4171 [hep-ph]} \BibitemShut {NoStop}%
\bibitem [{\citenamefont {Luque}\ \emph {et~al.}(2024)\citenamefont {Luque}, \citenamefont {Winkler},\ and\ \citenamefont {Linden}}]{luque2024cosmicraypropagationmodelselucidate}%
  \BibitemOpen
  \bibfield  {author} {\bibinfo {author} {\bibfnamefont {P.~D. L.~T.}\ \bibnamefont {Luque}}, \bibinfo {author} {\bibfnamefont {M.~W.}\ \bibnamefont {Winkler}},\ and\ \bibinfo {author} {\bibfnamefont {T.}~\bibnamefont {Linden}},\ }\href {https://arxiv.org/abs/2404.13114} {\bibinfo {title} {Cosmic-ray propagation models elucidate the prospects for antinuclei detection}} (\bibinfo {year} {2024}),\ \Eprint {https://arxiv.org/abs/2404.13114} {arXiv:2404.13114 [astro-ph.HE]} \BibitemShut {NoStop}%
\bibitem [{\citenamefont {Uzhinsky}\ \emph {et~al.}(2011)\citenamefont {Uzhinsky}, \citenamefont {Apostolakis}, \citenamefont {Galoyan}, \citenamefont {Folger}, \citenamefont {Grichine}, \citenamefont {Ivanchenko},\ and\ \citenamefont {Wright}}]{Uzhinsky:2011zz}%
  \BibitemOpen
  \bibfield  {author} {\bibinfo {author} {\bibfnamefont {V.}~\bibnamefont {Uzhinsky}}, \bibinfo {author} {\bibfnamefont {J.}~\bibnamefont {Apostolakis}}, \bibinfo {author} {\bibfnamefont {A.}~\bibnamefont {Galoyan}}, \bibinfo {author} {\bibfnamefont {G.}~\bibnamefont {Folger}}, \bibinfo {author} {\bibfnamefont {V.~M.}\ \bibnamefont {Grichine}}, \bibinfo {author} {\bibfnamefont {V.~N.}\ \bibnamefont {Ivanchenko}},\ and\ \bibinfo {author} {\bibfnamefont {D.~H.}\ \bibnamefont {Wright}},\ }\bibfield  {title} {\bibinfo {title} {{Antinucleus-nucleus cross sections implemented in Geant4}},\ }\href {https://doi.org/10.1016/j.physletb.2011.10.010} {\bibfield  {journal} {\bibinfo  {journal} {Phys. Lett. B}\ }\textbf {\bibinfo {volume} {705}},\ \bibinfo {pages} {235} (\bibinfo {year} {2011})}\BibitemShut {NoStop}%
\bibitem [{\citenamefont {Galoyan}\ \emph {et~al.}(2016)\citenamefont {Galoyan}, \citenamefont {Ribon},\ and\ \citenamefont {Uzhinsky}}]{Galoyan:2016huf}%
  \BibitemOpen
  \bibfield  {author} {\bibinfo {author} {\bibfnamefont {A.}~\bibnamefont {Galoyan}}, \bibinfo {author} {\bibfnamefont {A.}~\bibnamefont {Ribon}},\ and\ \bibinfo {author} {\bibfnamefont {V.}~\bibnamefont {Uzhinsky}},\ }\bibfield  {title} {\bibinfo {title} {{Dynamics of Anti-Proton\textendash{}Protons and Anti-Proton\textendash{}Nucleus Reactions}},\ }\href@noop {} {\bibfield  {journal} {\bibinfo  {journal} {Nucl. Theor.}\ }\textbf {\bibinfo {volume} {35}},\ \bibinfo {pages} {194} (\bibinfo {year} {2016})},\ \Eprint {https://arxiv.org/abs/1610.08341} {arXiv:1610.08341 [nucl-th]} \BibitemShut {NoStop}%
\bibitem [{\citenamefont {Allison}\ \emph {et~al.}(2016)\citenamefont {Allison} \emph {et~al.}}]{Allison:2016lfl}%
  \BibitemOpen
  \bibfield  {author} {\bibinfo {author} {\bibfnamefont {J.}~\bibnamefont {Allison}} \emph {et~al.},\ }\bibfield  {title} {\bibinfo {title} {{Recent developments in Geant4}},\ }\href {https://doi.org/10.1016/j.nima.2016.06.125} {\bibfield  {journal} {\bibinfo  {journal} {Nucl. Instrum. Meth. A}\ }\textbf {\bibinfo {volume} {835}},\ \bibinfo {pages} {186} (\bibinfo {year} {2016})}\BibitemShut {NoStop}%
\bibitem [{\citenamefont {Barashenkov}\ and\ \citenamefont {Toneev}(1972)}]{Barashenkov:1972zza}%
  \BibitemOpen
  \bibfield  {author} {\bibinfo {author} {\bibfnamefont {V.~S.}\ \bibnamefont {Barashenkov}}\ and\ \bibinfo {author} {\bibfnamefont {V.~D.}\ \bibnamefont {Toneev}},\ }\href@noop {} {\emph {\bibinfo {title} {{Interaction of High Energy Particles and Atomic Nuclei with Nuclei}}}}\ (\bibinfo  {publisher} {Atomizadt},\ \bibinfo {address} {Moscow},\ \bibinfo {year} {1972})\BibitemShut {NoStop}%
\bibitem [{\citenamefont {Golubeva}\ \emph {et~al.}(1988)\citenamefont {Golubeva}, \citenamefont {Ilinov}, \citenamefont {Botvina},\ and\ \citenamefont {Sobolevsky}}]{Golubeva:1988de}%
  \BibitemOpen
  \bibfield  {author} {\bibinfo {author} {\bibfnamefont {E.~S.}\ \bibnamefont {Golubeva}}, \bibinfo {author} {\bibfnamefont {A.~S.}\ \bibnamefont {Ilinov}}, \bibinfo {author} {\bibfnamefont {A.~S.}\ \bibnamefont {Botvina}},\ and\ \bibinfo {author} {\bibfnamefont {N.~M.}\ \bibnamefont {Sobolevsky}},\ }\bibfield  {title} {\bibinfo {title} {{Inelastic Interactions of Intermediate-energy Anti-nucleons With Nuclei}},\ }\href {https://doi.org/10.1016/0375-9474(88)90083-8} {\bibfield  {journal} {\bibinfo  {journal} {Nucl. Phys. A}\ }\textbf {\bibinfo {volume} {483}},\ \bibinfo {pages} {539} (\bibinfo {year} {1988})}\BibitemShut {NoStop}%
\bibitem [{\citenamefont {Barashenkov}\ \emph {et~al.}(1971)\citenamefont {Barashenkov}, \citenamefont {Ilinov},\ and\ \citenamefont {Toneev}}]{Barashenkov:1971vh}%
  \BibitemOpen
  \bibfield  {author} {\bibinfo {author} {\bibfnamefont {V.~S.}\ \bibnamefont {Barashenkov}}, \bibinfo {author} {\bibfnamefont {A.~S.}\ \bibnamefont {Ilinov}},\ and\ \bibinfo {author} {\bibfnamefont {V.~D.}\ \bibnamefont {Toneev}},\ }\bibfield  {title} {\bibinfo {title} {{Further development of the intranuclear cascade model}},\ }\href@noop {} {\bibfield  {journal} {\bibinfo  {journal} {Yad. Fiz.}\ }\textbf {\bibinfo {volume} {13}},\ \bibinfo {pages} {743} (\bibinfo {year} {1971})}\BibitemShut {NoStop}%
\bibitem [{\citenamefont {Golubeva}\ \emph {et~al.}(2019)\citenamefont {Golubeva}, \citenamefont {Barrow},\ and\ \citenamefont {Ladd}}]{Golubeva:2018mrz}%
  \BibitemOpen
  \bibfield  {author} {\bibinfo {author} {\bibfnamefont {E.~S.}\ \bibnamefont {Golubeva}}, \bibinfo {author} {\bibfnamefont {J.~L.}\ \bibnamefont {Barrow}},\ and\ \bibinfo {author} {\bibfnamefont {C.~G.}\ \bibnamefont {Ladd}},\ }\bibfield  {title} {\bibinfo {title} {{Model of $\bar n$ annihilation in experimental searches for $\bar n$ transformations}},\ }\href {https://doi.org/10.1103/PhysRevD.99.035002} {\bibfield  {journal} {\bibinfo  {journal} {Phys. Rev. D}\ }\textbf {\bibinfo {volume} {99}},\ \bibinfo {pages} {035002} (\bibinfo {year} {2019})},\ \Eprint {https://arxiv.org/abs/1804.10270} {arXiv:1804.10270 [hep-ex]} \BibitemShut {NoStop}%
\bibitem [{\citenamefont {Barashenkov}\ \emph {et~al.}(1973)\citenamefont {Barashenkov}, \citenamefont {Ilinov},\ and\ \citenamefont {Toneev}}]{Barashenkov:1973ew}%
  \BibitemOpen
  \bibfield  {author} {\bibinfo {author} {\bibfnamefont {V.~S.}\ \bibnamefont {Barashenkov}}, \bibinfo {author} {\bibfnamefont {A.~S.}\ \bibnamefont {Ilinov}},\ and\ \bibinfo {author} {\bibfnamefont {V.~D.}\ \bibnamefont {Toneev}},\ }\bibfield  {title} {\bibinfo {title} {{Interaction of relativistic deuterium and tritium nuclei with emulsion nuclei}},\ }\href@noop {} {\bibfield  {journal} {\bibinfo  {journal} {Acta Phys. Polon. B}\ }\textbf {\bibinfo {volume} {4}},\ \bibinfo {pages} {219} (\bibinfo {year} {1973})}\BibitemShut {NoStop}%
\bibitem [{\citenamefont {Barashenkov}(1993)}]{Barashenkov:1993zza}%
  \BibitemOpen
  \bibfield  {author} {\bibinfo {author} {\bibfnamefont {V.~S.}\ \bibnamefont {Barashenkov}},\ }\href@noop {} {\emph {\bibinfo {title} {{Cross Sections of Interaction of Particles and Nuclei with Nuclei}}}}\ (\bibinfo  {publisher} {JINR},\ \bibinfo {address} {Dubna},\ \bibinfo {year} {1993})\BibitemShut {NoStop}%
\bibitem [{\citenamefont {Paphomov}\ and\ \citenamefont {Sergeev}(2005)}]{Paph:2005kei}%
  \BibitemOpen
  \bibfield  {author} {\bibinfo {author} {\bibfnamefont {V.~E.}\ \bibnamefont {Paphomov}}\ and\ \bibinfo {author} {\bibfnamefont {V.~E.}\ \bibnamefont {Sergeev}},\ }\bibfield  {title} {\bibinfo {title} {{Application of the three-body model to describe the inelastic interaction of deuterons with nuclei}},\ }\href@noop {} {\bibfield  {journal} {\bibinfo  {journal} {Brief Reports on Physics of the FIAN}\ } (\bibinfo {year} {2005})}\BibitemShut {NoStop}%
\bibitem [{\citenamefont {Minomo}\ \emph {et~al.}(2017)\citenamefont {Minomo}, \citenamefont {Washiyama},\ and\ \citenamefont {Ogata}}]{Minomo:2016uto}%
  \BibitemOpen
  \bibfield  {author} {\bibinfo {author} {\bibfnamefont {K.}~\bibnamefont {Minomo}}, \bibinfo {author} {\bibfnamefont {K.}~\bibnamefont {Washiyama}},\ and\ \bibinfo {author} {\bibfnamefont {K.}~\bibnamefont {Ogata}},\ }\bibfield  {title} {\bibinfo {title} {{Deuteron\textendash{}nucleus total reaction cross sections up to 1 GeV}},\ }\href {https://doi.org/10.1080/00223131.2016.1213672} {\bibfield  {journal} {\bibinfo  {journal} {J. Nucl. Sci. Tech.}\ }\textbf {\bibinfo {volume} {54}},\ \bibinfo {pages} {127} (\bibinfo {year} {2017})},\ \Eprint {https://arxiv.org/abs/1608.05503} {arXiv:1608.05503 [nucl-th]} \BibitemShut {NoStop}%
\bibitem [{\citenamefont {Kharzeev}\ and\ \citenamefont {Khrapov}(1986)}]{kharzeev1986kulonovskii2377}%
  \BibitemOpen
  \bibfield  {author} {\bibinfo {author} {\bibfnamefont {D.}~\bibnamefont {Kharzeev}}\ and\ \bibinfo {author} {\bibfnamefont {A.~A.}\ \bibnamefont {Khrapov}},\ }\bibfield  {title} {\bibinfo {title} {Coulomb break-up of an antideuteron atom},\ }\href@noop {} {\bibfield  {journal} {\bibinfo  {journal} {Moscow University Physics Bulletin}\ }\textbf {\bibinfo {volume} {4}},\ \bibinfo {pages} {27} (\bibinfo {year} {1986})}\BibitemShut {NoStop}%
\bibitem [{\citenamefont {Paphomov}\ and\ \citenamefont {Sergeev}(2002)}]{Paph:2002kei}%
  \BibitemOpen
  \bibfield  {author} {\bibinfo {author} {\bibfnamefont {V.~E.}\ \bibnamefont {Paphomov}}\ and\ \bibinfo {author} {\bibfnamefont {V.~E.}\ \bibnamefont {Sergeev}},\ }\bibfield  {title} {\bibinfo {title} {{The contribution of diffraction splitting of deuterons to the total cross sections of reactions with light nuclei}},\ }\href@noop {} {\bibfield  {journal} {\bibinfo  {journal} {Brief Reports on Physics of the FIAN}\ } (\bibinfo {year} {2002})}\BibitemShut {NoStop}%
\bibitem [{\citenamefont {Maurin}\ \emph {et~al.}(2025)\citenamefont {Maurin}, \citenamefont {Audouin}, \citenamefont {Berti}, \citenamefont {Coppin}, \citenamefont {Mauro}, \citenamefont {von Doetinchem}, \citenamefont {Donato}, \citenamefont {Evoli}, \citenamefont {Génolini}, \citenamefont {Ghosh}, \citenamefont {Leya}, \citenamefont {Losekamm}, \citenamefont {Mariani}, \citenamefont {Norbury}, \citenamefont {Orusa}, \citenamefont {Paniccia}, \citenamefont {Poeschl}, \citenamefont {Serpico}, \citenamefont {Tykhonov}, \citenamefont {Unger}, \citenamefont {Vanstalle}, \citenamefont {Zhao}, \citenamefont {Boncioli}, \citenamefont {Chiosso}, \citenamefont {Giordano}, \citenamefont {Coral}, \citenamefont {Graziani}, \citenamefont {Lucarelli}, \citenamefont {Maestro}, \citenamefont {Mahlein}, \citenamefont {Morejon}, \citenamefont {Ocampo-Peleteiro}, \citenamefont {Oliva}, \citenamefont {Pierog},\ and\ \citenamefont {Šerkšnytė}}]{maurin2025precisioncrosssectionsadvancingcosmicray}%
  \BibitemOpen
  \bibfield  {author} {\bibinfo {author} {\bibfnamefont {D.}~\bibnamefont {Maurin}}, \bibinfo {author} {\bibfnamefont {L.}~\bibnamefont {Audouin}}, \bibinfo {author} {\bibfnamefont {E.}~\bibnamefont {Berti}}, \bibinfo {author} {\bibfnamefont {P.}~\bibnamefont {Coppin}}, \bibinfo {author} {\bibfnamefont {M.~D.}\ \bibnamefont {Mauro}}, \bibinfo {author} {\bibfnamefont {P.}~\bibnamefont {von Doetinchem}}, \bibinfo {author} {\bibfnamefont {F.}~\bibnamefont {Donato}}, \bibinfo {author} {\bibfnamefont {C.}~\bibnamefont {Evoli}}, \bibinfo {author} {\bibfnamefont {Y.}~\bibnamefont {Génolini}}, \bibinfo {author} {\bibfnamefont {P.}~\bibnamefont {Ghosh}}, \bibinfo {author} {\bibfnamefont {I.}~\bibnamefont {Leya}}, \bibinfo {author} {\bibfnamefont {M.~J.}\ \bibnamefont {Losekamm}}, \bibinfo {author} {\bibfnamefont {S.}~\bibnamefont {Mariani}}, \bibinfo {author} {\bibfnamefont {J.~W.}\ \bibnamefont {Norbury}}, \bibinfo {author} {\bibfnamefont {L.}~\bibnamefont {Orusa}}, \bibinfo {author} {\bibfnamefont {M.}~\bibnamefont
  {Paniccia}}, \bibinfo {author} {\bibfnamefont {T.}~\bibnamefont {Poeschl}}, \bibinfo {author} {\bibfnamefont {P.~D.}\ \bibnamefont {Serpico}}, \bibinfo {author} {\bibfnamefont {A.}~\bibnamefont {Tykhonov}}, \bibinfo {author} {\bibfnamefont {M.}~\bibnamefont {Unger}}, \bibinfo {author} {\bibfnamefont {M.}~\bibnamefont {Vanstalle}}, \bibinfo {author} {\bibfnamefont {M.~J.}\ \bibnamefont {Zhao}}, \bibinfo {author} {\bibfnamefont {D.}~\bibnamefont {Boncioli}}, \bibinfo {author} {\bibfnamefont {M.}~\bibnamefont {Chiosso}}, \bibinfo {author} {\bibfnamefont {D.}~\bibnamefont {Giordano}}, \bibinfo {author} {\bibfnamefont {D.~M.~G.}\ \bibnamefont {Coral}}, \bibinfo {author} {\bibfnamefont {G.}~\bibnamefont {Graziani}}, \bibinfo {author} {\bibfnamefont {C.}~\bibnamefont {Lucarelli}}, \bibinfo {author} {\bibfnamefont {P.}~\bibnamefont {Maestro}}, \bibinfo {author} {\bibfnamefont {M.}~\bibnamefont {Mahlein}}, \bibinfo {author} {\bibfnamefont {L.}~\bibnamefont {Morejon}}, \bibinfo {author} {\bibfnamefont
  {J.}~\bibnamefont {Ocampo-Peleteiro}}, \bibinfo {author} {\bibfnamefont {A.}~\bibnamefont {Oliva}}, \bibinfo {author} {\bibfnamefont {T.}~\bibnamefont {Pierog}},\ and\ \bibinfo {author} {\bibfnamefont {L.}~\bibnamefont {Šerkšnytė}},\ }\href {https://arxiv.org/abs/2503.16173} {\bibinfo {title} {Precision cross-sections for advancing cosmic-ray physics and other applications: a comprehensive programme for the next decade}} (\bibinfo {year} {2025}),\ \Eprint {https://arxiv.org/abs/2503.16173} {arXiv:2503.16173 [astro-ph.HE]} \BibitemShut {NoStop}%
\end{thebibliography}%

\end{document}